\documentclass[
superscriptaddress,
nofootinbib,
 amsmath,amssymb,
 aps, prd,
]{revtex4-1}

\usepackage{graphicx}
\usepackage{dcolumn}
\usepackage{bm}
\usepackage{hyperref}
\usepackage{slashed}
\usepackage[dvipsnames]{xcolor}
\usepackage{float}

\begin{document}

\preprint{CPHT-RR040.062020}

\title{Diffractive two-meson electroproduction with a nucleon and deuteron target}

\author{W. Cosyn}
\email[ E-mail: ]{wcosyn@fiu.edu}
\affiliation{%
 Department of Physics, Florida International University,  Miami, Florida 3199, USA
}%
\affiliation{
 Department of Physics and Astronomy, Ghent University, B9000 Ghent, Belgium
}

\author{ B.~Pire}
\email[ E-mail: ]{bernard.pire@polytechnique.edu}
\affiliation{ Centre de Physique Th\'eorique, CNRS, \'Ecole Polytechnique, I. P. Paris,
  91128 Palaiseau,     France  }
\author{ L.~Szymanowski}
\email[ E-mail: ]{Lech.Szymanowski@ncbj.gov.pl}
\affiliation{National Centre for Nuclear Research (NCBJ), Pasteura 7, 02-093 Warsaw, Poland}

\date{\today}

\begin{abstract}
The diffractive electro- or photo-production of two mesons separated by a large rapidity gap gives access to generalized parton distributions (GPDs) in a very specific way. First, these reactions allow to easily access the chiral-odd transversity quark GPDs by selecting one of the produced vector mesons to be transversely polarized. Second, they are only sensitive to the so-called ERBL region where GPDs are not much constrained by forward quark distributions. Third, the skewness parameter  $\xi$ is not related to the Bjorken $x_\text{Bj}$ variable, but to the size of the rapidity gap. We analyze different channels ($\rho_L^0\,\rho_{L/T}, \rho^0_L\,\omega_{L/T}$ and $\rho^0_L\,\pi$ production) on nucleon and deuteron targets.  The analysis is performed in the kinematical domain where a large momentum transfer from the photon to the diffractively produced  vector meson  introduces a hard scale (the virtuality of the exchanged hard Pomeron).  This enables the description of the hadronic part of the process in the framework of collinear factorization of GPDs.  We show that the unpolarized cross sections depend very much on the parameterizations of both chiral-even and chiral-odd quark distributions of the nucleon, as well as on the shape of the meson distribution amplitudes. The rates are shown to be in the range of the capacities of a future electron-ion collider.   
\end{abstract}

\maketitle


\section{Introduction}
\label{sec:intro}
Diffractive events are known to constitute a large part of the cross section in high-energy scattering. Their understanding in the framework of quantum chromodynamics (QCD) is based on the concept of Pomeron exchange between two subprocesses; in the Regge inspired $k_T-$factorization approach valid at high energy and at the leading order, the scattering amplitude is written in terms of  two impact factors with at least two reggeized gluon exchange in the $t-$channel. 

Provided on the one hand that a hard scale allows the use of perturbative QCD, and on the other hand that the kinematical regimes are in the so-called generalized Bjorken region~\cite{Mueller:1998fv}, the impact factors can be calculated in the collinear factorization framework, with a perturbatively calculable coefficient function convoluted with  distribution amplitudes (DAs) and generalized parton distributions (GPDs)  encoding their long distance parts. 

\begin{figure}[ht]
    \centering
\includegraphics[width=0.5\textwidth]{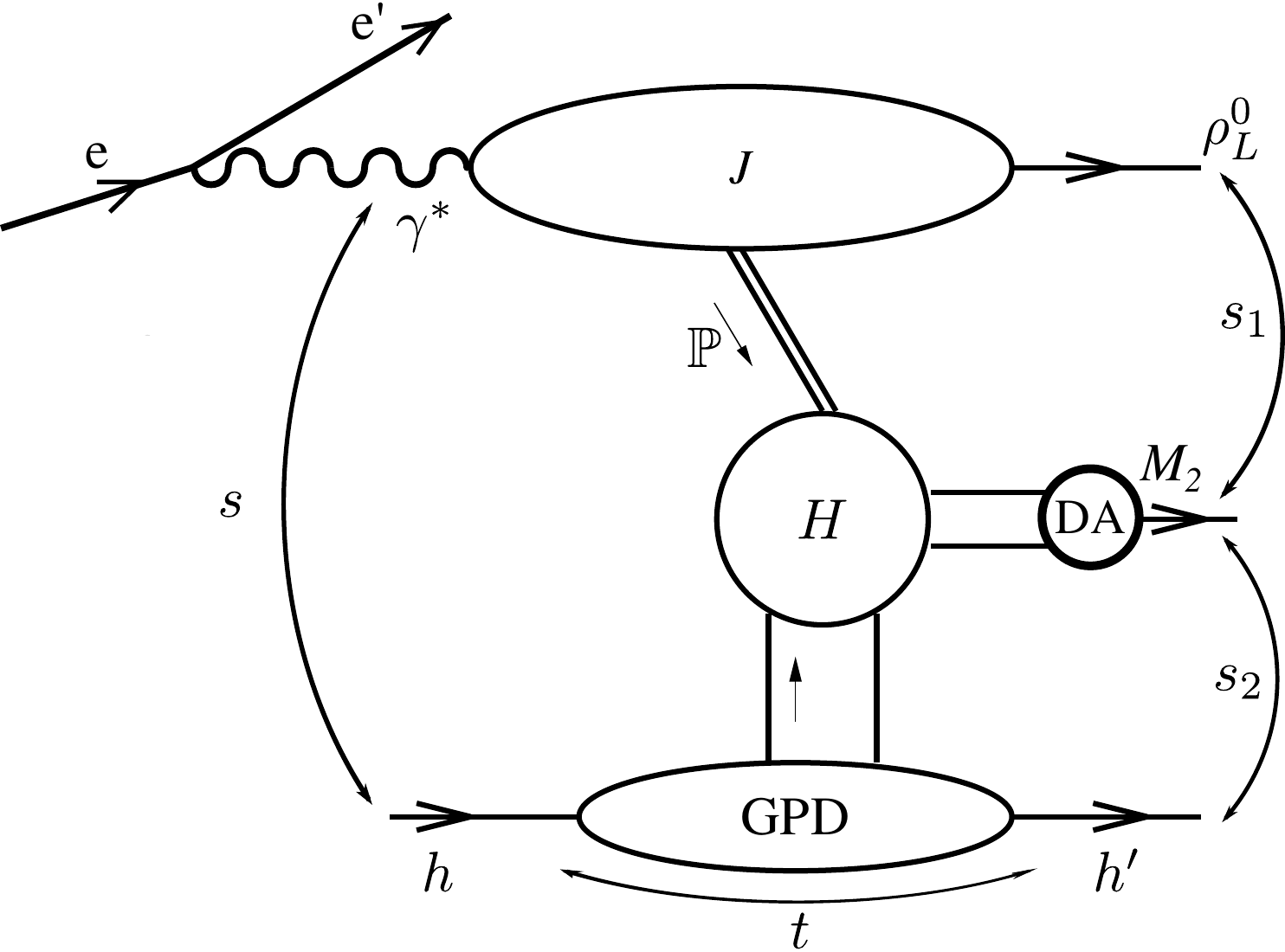}
\caption{The diffractive process $e + h \to e' + \rho_L^0 + M_2 + h'$ amplitude in the hybrid factorization approach.  $H$ represents the hard part of the scattering, $J$ the photon impact factor and the GPD [DA] vertices represent the nucleon [meson] quark-quark hadron correlators that are appropriately  parameterized.  See Sec.~\ref{subsec:kinematics} for the definition of the kinematic invariants $s,s_1,s_2,t$.}
    \label{fig:process}
\end{figure}
This hybrid factorization picture has been shown to be valid, at least to the leading order, in the pioneering study \cite{Ivanov:2002jj} where the exclusive process 
\begin{equation}
 e + N \to e' + \rho_L^0 + \rho^+ + N'\,,   
\end{equation}
has been studied in the kinematical regime where the two $\rho$ mesons are separated by a large rapidity gap and the hard scale is the virtuality of the hard Pomeron, see Fig.\ref{fig:process}.  This hard scale ensures the small-sized configuration in the top part of the diagram ($J$) and the separation of short-distance ($H$) and long-distance (GPD, DA) dynamics in the bottom part of Fig.~\ref{fig:process}.  In Ref.\cite{Enberg:2006he}, this same reaction has been discussed with a particular emphasis on its sensitivity to the elusive transversity chiral-odd GPDs~\cite{Diehl:2001pm}. For this purpose, a model for one transversity GPD ($H_T(x,\xi,t)$) was developed.

More than one decade later, these transversity quark distributions are less mysterious although still largely undetermined~\cite{Gockeler:2006zu, Ahmad:2008hp,Goloskokov:2009ia,Goloskokov:2011rd,Goldstein:2012az,Schweitzer:2016jmd,Cosyn:2018rdm}, mostly because of their absence in leading twist amplitudes for simple processes \cite{Diehl:1998pd, Collins:1999un}. Moreover, there is the vigorous rise of international interest for the electron-ion collider (EIC) project~\cite{Boer:2011fh,Accardi:2012qut} at Brookhaven National Laboratory.  The EIC, equipped with suitable forward detectors, opens new opportunities to measure these electroproduction cross sections at  energies high enough to justify the formalism we will sketch in Section \ref{Formalism}.  In the further future, other electron-ion colliders such as the Large Hadron electron Collider (LHeC)~\cite{AbelleiraFernandez:2012cc} or the Electron-ion collider in China (EicC) project \cite{Chen:2018wyz} may offer additional possibilities.  It is thus deemed appropriate to revisit and enlarge the phenomenology of two-meson electroproduction in the hard diffractive regime, adding the coherent deuteron target case which has recently attracted much attention together with other light nucleus studies \cite{Dupre:2015jha, Fucini:2018gso}. This is the goal of this study, where we shall focus on the following reactions \footnote{Everywhere in this paper, the first meson $M_1$ in a pair $M_1 M_2$ is diffractively produced while the second one $M_2$ comes from the reaction ${\mathbb P} h \to M_2 h'$, as depicted in Fig.\ref{fig:process}. In a leading twist calculation, the first meson $M_1$ cannot be a transversely polarized vector meson.}:
\begin{eqnarray}
 e + N \to e'+  \rho^0_L + \rho^0_L + N'\,~~&,&~~ 
 e + N \to e' + \rho^0_L + \rho^0_T+  N'\,,\nonumber\\
 e +  N \to e' + \rho^0_L + \omega_L + N'\, ~~&,&~~ 
 e + N \to e' + \rho^0_L + \omega_T + N'\,,\\
  e + N &\to& e' + \rho^0_L + \pi^0 + N'\,,\nonumber
\end{eqnarray}
on a nucleon target, and
\begin{eqnarray}
 e + D \to e' + \rho^0_L + \omega_L + D'\,~~&,&~~
 e + D \to e' + \rho^0_L + \omega_T + D'\,,
\end{eqnarray}
on a deuteron target in coherent reactions.  Here, the subscripts $L/T$ refer to the longitudinal or transverse polarization of the vector mesons.

Let us now summarize the characteristic features of these reactions in the chosen kinematical domain and in the leading order approximation used to calculate their amplitudes~\cite{Ivanov:2002jj,Enberg:2006he}:
\begin{itemize}
    \item The cross sections are independent of the $\gamma^{(*)} N$ or $\gamma^{(*)} D$ squared invariant mass $s$, and their rates are quite large in the high energy domain.
    \item Contrarily to the usual deeply virtual Compton scattering (DVCS) case, the skewness variable (see Eq.~(\ref{eq:t_and_xi})) is not related to Bjorken variable $x_\text{Bj}$ but to the ratio $z= \frac{s_1}{s}$ through the relation $\xi \approx \frac{z}{2-z}$.  Here $s_1$ and $s$ are the invariant squared masses of the final two-meson system, and photon-hadron system respectively; see Fig.~\ref{fig:process} and Eq. (\ref{eq:s1}) below.
    \item The amplitudes depend on the so-called Efremov-Radyushkin-Brodsky-Lepage (ERBL) region of GPDs ($-\xi < x < \xi$) where GPDs are particularly unrestricted, in particular because the positivity constraints \cite{Pire:1998nw} which relates them to usual quark distributions do not apply.\footnote{This property is also present in the diffractive DVCS reaction \cite{Pire:2019hos}.}
    \item The amplitudes do not depend on gluon GPDs, due to the $C$-parity of the meson ($\rho^0,\omega$) or isospin ($\rho^0,\pi$).
    \item The amplitudes for $\rho\rho$ or $\rho\,\omega$ production depend on the charge conjugation odd GPDs.\footnote{This property is also present in the diphoton electroproduction case \cite{Pedrak:2017cpp, Pedrak:2020mfm}.}  For the nucleon these are the chiral-even GPDs 
    \begin{equation}
    H^-(x,\xi,t) = \tfrac{1}{2}[H(x,\xi,t) + H(-x,\xi,t)]   \,,
    \end{equation}
    together with the corresponding $E^-(x,\xi,t)$ and the  chiral-odd GPDs $H_T^-(x,\xi,t)$, $E_T^-(x,\xi,t)$, $\widetilde{H}_T^-(x,\xi,t)$, and $\widetilde{E}_T^-(x,\xi,t)$. These charge conjugation odd GPDs do not have $D-$terms~\cite{Polyakov:1999gs}.
    \item The amplitudes for $\rho \,\pi$ production depend on the charge conjugation even GPDs 
    \begin{equation}
     \widetilde H^+(x,\xi,t) = \tfrac{1}{2}[\widetilde H(x,\xi,t) + \widetilde H(-x,\xi,t)]   
    \end{equation}
    and the corresponding $\widetilde E^+(x,\xi,t)$ GPD. 
    \item The isoscalar nature of the deuteron selects the $\rho\,\omega$ channel.
\end{itemize}
The paper is organized as follows. In Section II, we describe precisely the kinematics we are interested in and recall the formalism used for calculating the scattering amplitude at high energy. In Section III, we review the parameterizations of the various non-perturbative inputs needed to perform the cross section estimates, namely the distribution amplitudes (DAs) of the produced mesons and the generalized parton distributions (GPDs) of the nucleon and deuteron, the latter calculated in a convolution model. Section IV shows our predictions for the different channels and discusses the interesting sensitivity of some observables to the yet poorly known GPDs. Section V gives our conclusions and comments on the feasibility of the measurement of the discussed processes in the EIC project.

\section{Formalism}
\label{sec:formalism}
\subsection{Kinematics}
\label{subsec:kinematics}
In our process
\begin{equation}
\label{eq:photoreaction}
    \gamma^* (q) + h(p_h) \rightarrow \rho^0_L(q_\rho) + M_2(p_{M_2}) + h'(p'_h) \,,
\end{equation}
we decompose the four-vectors as follows.
We introduce two light-like Sudakov vectors $p$ and $n$ such that 
 $p=\tfrac{1}{2}(p_h +p'_h)$ at $-t=(-t)_\text{min}=\tfrac{4\xi^2M^2}{1-\xi^2}$ (with $M$ the mass of $h$) and $n=q$ when $q^2=0$; see Eq.~(\ref{Sud}). We also introduce the auxiliary variable
${\cal S}=2(pn)$,  related to the total  center-of-mass energy squared of the 
$\gamma^*p-$system, $s=(q+p_h)^2$, as $s+Q^2=(1+\xi){\cal S}$.  The momentum transfer $t$ to the target hadron and skewness $\xi$ are
\begin{align}\label{eq:t_and_xi}
&t=\Delta^2 \quad [\Delta \equiv p'_h-p_h]\,, &\xi = -\frac{(\Delta n)}{2(pn)}\,.
\end{align}

In the collinear approximation  which is suitable for the calculation of the hard coefficient, the momenta are parameterized as follows (neglecting hadron masses and keeping $-t=(-t)_\text{min}$):
\begin{eqnarray}
\label{Sud}
 && q^\mu = n^\mu  - \frac{Q^2}{{\cal S}}p^\mu \;, 
\nonumber \\
&& p_h^\mu = (1+\xi)p^\mu\,,\;\;\;\;\;({p_h'})^{\;\mu} = (1-\xi)p^\mu\;,\nonumber\\
&& q_\rho^\mu = \alpha n^\mu +
\frac{\bm q_\rho ^{\,2}}{\alpha {\cal S}}p^\mu + q_{\rho T}^\mu \;,\;\;\;\;\;q_{\rho T}^{2} =
-\bm q_\rho^{\,2}\; \qquad [(n\,q_{\rho T})=(p\,q_{\rho T})=0],
\nonumber \\
&& p_{M_2}^\mu = \bar \alpha n^\mu 
+ \frac{\bm q_{\rho}^{\,2}}{\bar \alpha {\cal S}}p^\mu
- q_{\rho T}^\mu \;,\;\;\;\;\;\bar \alpha \equiv 1-\alpha\;,
\end{eqnarray}
where $Q^2=-q^2$ is the photon virtuality. The large invariant squared mass of the two produced mesons is
\begin{equation}\label{eq:s1}
    s_1 = (q_\rho + p_{M_2})^2 \approx \frac{\bm q_\rho^{\,2}}{\alpha \bar \alpha}\,,
\end{equation}
and
\begin{equation}
    s_2 = (p_{M_2} + p'_h)^2 \approx \bar \alpha \frac{1-\xi}{1+\xi} (s+Q^2)
\end{equation}
is the squared energy of the Pomeron-nucleon reaction.
The kinematical regime with a large rapidity gap between the two mesons in the final state is
obtained by demanding that $s_1$ be very large,  of the order of 
$s$,
whereas $s_2$ is kept of the order of $\bm q_\rho^{2}$, quite smaller than $s_1$ but however
large enough to justify the use
of  perturbation theory in the
collinear subprocess ${\mathbb P} h \to M_2 h'$ and the application of
the GPD framework.

In terms of the longitudinal fraction $\alpha$ the limit
with a large rapidity gap corresponds
to taking the limits
\begin{equation}
 \label{alphagap}
\alpha \to 1\,,\;\;\;\;\;\bar \alpha s_1 \to \bm q_\rho^{\,2}\,.   
\end{equation}
The hard Pomeron virtuality (which as we remind provides the hard scale) in this limit corresponds to $-\bm{q}_\rho^{\,2}$.
Skewness $\xi$ can be written in
terms of the invariant mass $s_1$ of the two mesons as 
\begin{equation}
 \xi \approx \frac{s_1+Q^2}{2s-s_1+Q^2} \approx \frac{s_1}{2s-s_1}\,,
\label{s1}   
\end{equation}
which shows that at large energy, our process probes large values of $\xi$, for instance, $\xi \approx 0.5$ ($0.33$) when $s_1=2s/3$ ($s_1=s/2$). 

\subsection{Summary of formalism}
\label{Formalism}
Let us recall the results of Ref.~\cite{Ivanov:2002jj} for the scattering amplitudes. 
The amplitude for the production of two longitudinal $\rho_L^0$ mesons reads:
\begin{eqnarray}
\label{CEN}
{\cal M}^{\gamma^{(*)}_{L/T}\,h\,\to \rho_L^0\, \rho^0_L\,h'}&=&
i 16\pi^2   s \alpha_s \frac{f_{\rho}^L}{\sqrt{2}} \xi
\frac{C_F}{N_c\,(\bm q_\rho^{\,2})^2}
\nonumber \\
& \times & \int\limits_0^1
\frac{\;\mathrm{d}u\;\phi_\parallel(u)}{ \,u^2 \bar u^2 }
 J^{\gamma^{(*)}_{L/T} \to \rho^0_L}(u\bm{q}_\rho,\bar u\bm{q}_\rho)
 \int_{-1}^1 \mathrm{d}x \, \delta(x-\xi(2u-1)) \, V^{u-d}_{\lambda^\prime\lambda}(x,\xi,t) \,,
 \end{eqnarray}
with $\bar u = 1-u$, $N_c = 3, C_F=4/3$ and $\alpha_s$ the strong coupling constant.  For the longitudinal photon case, the impact
factor  $J^{\gamma^{*}_L \to \rho^0_L}$ is written (with $\bar z = 1-z$)  as
\begin{equation}
\label{ifgamma}
J^{\gamma^{(*)}_L \to \rho^0_L}(\bm k_1,\bm k_2 = \bm{q}_\rho-\bm{k}_1)=  -   f_\rho^L \frac{e
\alpha_s 2\pi
  Q}{N_c\sqrt{2}} \int\limits_0^1 \mathrm{d}z\;z\bar z 
\phi_\parallel(z)P(\bm k_1,\bm k_2)\;,
\end{equation}
with \footnote{In all our estimates, we put the quark mass $m_q=0$. See however the discussion in subsection IV-C.}
\begin{eqnarray}
\label{P}
P(\bm k_1,\bm k_2=\bm q_\rho-\bm k_1)=&&\frac{1}{z^2\bm q_\rho^{\,2}+m_q^2 +Q^2z\bar z} +
\frac{1}{{\bar z}^2\bm q_\rho^{\,2}+m_q^2 +Q^2z\bar z} \nonumber \\
&&-\frac{1}{(\bm k_1-z\bm q_\rho\,)^2+m_q^2 +Q^2z\bar z} -
\frac{1}{(\bm k_1-\bar z \bm q_\rho\,)^2+m_q^2 +Q^2z\bar z}\;.
\end{eqnarray}
In Eqs.~(\ref{CEN}) and (\ref{ifgamma}), the longitudinally polarized $\rho-$meson distribution amplitude (DA) $\phi_\parallel(z)$ and its normalization decay constant $f^L_\rho$ are defined by the matrix element~\cite{Ball:1996tb} 
\begin{equation}
\label{phi||}
\langle 0 | \bar u(0) \gamma^\mu u(y)|\rho^0_L(q_\rho)\rangle 
= q_\rho^\mu \frac{f_{\rho}^0}{\sqrt{2}} \int\limits_0^1\mathrm{d}z\;e^{-iz(q_\rho y)}\phi_\parallel(z)\;.
\end{equation}
In Eq.~(\ref{CEN}), the leading twist quark-hadron correlator 
\begin{equation}
V^q_{\lambda^\prime\lambda}(x,\xi,t)=\int_{-\infty}^\infty \frac{\mathrm{d}\kappa}{2\pi}
    \,e^{2ix(p  n)\kappa}\,\langle h'(p_h^\prime, \lambda^\prime) |\,
    \bar{q}(-n\kappa)
    \slashed{n}
    q(n\kappa)\,
    | h(p_h, \lambda) \rangle     
    \label{eq:GPDV}
\end{equation}
encodes the hadron long-distance structure through the chiral-even GPDs. Here $\lambda$ ($\lambda'$) refers to the spin quantum numbers of $h$ ($h'$).  In both Eqs.~(\ref{phi||}) and (\ref{eq:GPDV}) and similar matrix elements below a Wilson line is implicit.  At $-t=(-t)_\text{min}$, the matrix elements for the case of the nucleon are~\cite{Diehl:2003ny}  
\begin{equation}
\label{eq:corrV}
 V^q_{\lambda^\prime\lambda}(x,\xi,t) = \delta_{(\lambda',\lambda)}\sqrt{1-\xi^2} \left[  H^q(x,\xi,t) - \frac{\xi^2}{1-\xi^2}E^q(x,\xi,t)\right]\,.
\end{equation}
Equivalent deuteron expressions (ommitted here for length reasons) can be found in the appendix of Ref.~\cite{Cano:2003ju}.

For $\gamma^{(*)}$ transversely polarized ($\bm{\varepsilon}$ being its polarization vector), 
$J^{\gamma^{(*)}_T \to \rho^0_L}$ reads
\begin{equation}
\label{TL}
J^{\gamma^{(*)}_T \to \rho^0_L}(\bm k_1,\bm k_2=\bm{q}_\rho-\bm k_1)=
-\frac{e\,\alpha_s\,\pi\,f_\rho^L}{\sqrt{2}\,N_c}\;\int\limits_0^1\,\mathrm{d}z\,(2z-1)\,
\phi_\parallel(z)\,\left[ \bm{\varepsilon}\,\cdot\bm Q_P(\bm{k}_1,\bm{k}_2) \right]\;,
\end{equation}
with 
\begin{eqnarray}
\label{Q}
&&\bm Q_P(\bm k_1,\bm k_2=\bm q_\rho-\bm k_1)=
\frac{z\,\bm q_\rho}{z^2\,\bm q_\rho^2+Q^2\,z\,\bar z +m_q^2}
- \frac{\bar z\,\bm q_\rho}{\bar z^2\,\bm q_\rho^2+Q^2\,z\,\bar z +m_q^2}
 \\
&&\hspace{3cm}+\frac{\bm k_1 - z\,\bm q_\rho}{(\bm k_1 - z\,\bm q_\rho)^2+Q^2\,z\,\bar z +m_q^2}
-\frac{\bm k_1 - \bar z\,\bm q_\rho}{(\bm k_1 - \bar z\,\bm q_\rho)^2+Q^2\,z\,\bar z +m_q^2}\;.
\nonumber
\end{eqnarray}

The amplitude for the production of $\rho^0_L \omega_L$ is given by Eq.~(\ref{CEN}) with $f^L_\rho \rightarrow f^L_\omega$ and $V^{u-d}_{\lambda^\prime\lambda} \rightarrow V^{u+d}_{\lambda^\prime\lambda}$.
The amplitude for the production of $\rho^0_L \pi^0$ is given by Eq.~(\ref{CEN}) with  $f^L_\rho \rightarrow f_\pi$ and $V^{u-d}_{\lambda^\prime\lambda} \rightarrow \widetilde A^{u-d}_{\lambda^\prime\lambda}(x,\xi,t)$, where
\begin{equation}
      \widetilde{A}^q_{\lambda^\prime\lambda}(x,\xi,t)
    =
    \int_{-\infty}^\infty \frac{\mathrm{d}\kappa}{2\pi}
    \, e^{2ix(p  n)\kappa}\,
    \langle h'(p_h^\prime, \lambda^\prime) |\,
    \bar{q}(-n\kappa)
    \slashed{n} \gamma_5
    q(n\kappa)\,
    | h(p_h, \lambda) \rangle 
\end{equation}
is encoded through the leading twist axial GPDs. We have at $-t=(-t)_\text{min}$ for the nucleon~\cite{Diehl:2003ny}
\begin{equation}
\label{eq:corrA}
 \widetilde{A}^{u-d}_{\lambda^\prime\lambda}(x,\xi,t) = \delta_{(\lambda',\lambda)}\lambda \sqrt{1-\xi^2} \left[  \widetilde{H}^{u-d}(x,\xi,t) - \frac{\xi^2}{1-\xi^2}\widetilde{E}^{u-d}(x,\xi,t)\right]\,.
\end{equation}
For the equivalent deuteron expressions, we refer again to the appendix of Ref.~\cite{Cano:2003ju}.

In the case of the transversely polarized vector meson production, one obtains the amplitude~\cite{Ivanov:2002jj}
\begin{eqnarray}
\label{CON}
&&{\cal M}^{\gamma^{(*)}_{L/T} \,h\,\to \rho_L^0\, \rho^0_T\,h'}=
 -\,16\pi^2 s \alpha_s \frac{f_{\rho}^T}{\sqrt{2}} \xi
\frac{C_F}{N_c\,(\bm q_\rho^{\,2})^2}
\nonumber \\
&&\times\int\limits_0^1
\frac{\;\mathrm{d}u\;\phi_\perp(u)}{ \,u^2 \bar u^2 }
 J^{\gamma^{(*)}_{L/T} \to \rho^0_L}(u\bm p,\bar u\bm p)\int \mathrm{d}x \, \delta(x-\xi(2u-1)) \, \left[ \bm \epsilon_{\rho T} \cdot \bm T^{u-d}_{\lambda^\prime\lambda} (x,\xi,t) \right]\,,
\end{eqnarray}
where $\bm{\epsilon}_{\rho T}$ is the polarization three-vector
 of the produced transversely polarized vector meson and the transversity leading twist quark-hadron correlator 
\begin{equation}
    \bm{T}^{q\,i}_{\lambda^\prime\lambda}
    =
    \int_{-\infty}^\infty \frac{\mathrm{d}\kappa}{2\pi}\,
    e^{2ix(p  n)\kappa}\,
    \langle h'(p_h^\prime, \lambda^\prime) |\,
    \bar{q}(-n\kappa)
    (n_\mu \sigma^{\mu i})
    q(n\kappa)\,
    | h(p_h, \lambda) \rangle
\end{equation}
 is parameterized with the chiral-odd
 GPDs.  At $-t=(-t)_\text{min}$ we have for the nucleon~\cite{Diehl:2001pm}
\begin{equation}
\label{eq:corrT}
    \bm \epsilon_{\rho T} \cdot \bm T^{u-d}_{\lambda^\prime\lambda} (x,\xi,t) = \delta_{(\lambda',-\lambda)}\sin \theta \sqrt{1-\xi^2}\left[H^{u-d}_T(x,\xi,t)-\frac{\xi^2}{1-\xi^2}(E^{u-d}_T-\widetilde{E}^{u-d}_T) \right]\,,
\end{equation}
where $\theta$ is the angle between the transverse 
polarization vector of
the target nucleon $\bm{s}_T$ and $\bm \epsilon_{\rho T}$.  Equivalent deuteron expressions can be found in the appendix of Ref.~\cite{Cosyn:2018rdm}.
The corresponding amplitude for the  $\rho^0_L \omega_T$ case is given by Eq.~(\ref{CON}) with $f^T_\rho \rightarrow f^T_\omega$ and $\bm T^{u-d}_{\lambda'\lambda} \rightarrow T^{u+d}_{\lambda'\lambda}$. In Eq.~(\ref{CON}), $J^{\gamma^{(*)}_{L/T} \to \rho^0_L}$ are the same impact factors 
defined in
Eqs.~(\ref{ifgamma}) and (\ref{TL}).  The transversely polarized vector meson DA $\phi_\perp(z)$ together with its normalization decay constant $f^T$  reads \cite{Ball:1996tb}:
\begin{equation}
\label{phiT}
\langle 0 \mid \bar u(0) \sigma^{\mu \nu} u(x)\mid \rho_T(p_\rho,\epsilon_{\rho T})
\rangle =i \frac{f_\rho^T}{\sqrt{2}} \left( \epsilon^{\mu}_{\rho T} p_\rho^{\nu} -
p_\rho^{\mu}\epsilon^{\nu}_{\rho T}
\right)
\int\limits_0^1 \mathrm{d}z e^{-iz(p_\rho x)}\;\phi_\perp(z)\;. 
\end{equation}
 We shall show below the results for  the $\theta$-averaged cross section\footnote{This results in $\sin^2\theta \rightarrow \frac{1}{2}$ in the cross section.}  i.e. for the unpolarized target case.  The $\sin\theta$-modulation of the amplitude could of course be confirmed experimentally in an experiment with a transversely polarized target and data binned  differential in a $\theta$-dependent variable, but this is beyond the scope of this study.
 
The differential virtual photon cross section of Eq.~(\ref{eq:photoreaction}) at $-t = (-t)_\text{min}$ has the following form~\cite{Enberg:2006he}
\begin{equation}\label{eq:cross}
 \frac{\mathrm{d}\sigma}{\mathrm{d} \bm q_\rho^2 \mathrm{d}|t| \mathrm{d}\xi} = \frac{1}{256\pi^3\xi(1+\xi)s^2} \sum_{\text{spins}} |\mathcal{M}|^2\,,
\end{equation}
where we integrated over azimuthal angles of the $\rho^0$ and $h'$, and $\sum_{\text{spins}}$ denotes averaging and summing over the relevant spin degrees ($h$, $h'$, $M_2$); in this work we consider the completely unpolarized hadron ($h,h'$) case but longitudinal and transverse polarization of $M_2$ separately, using Eq.~(\ref{CEN}) and (\ref{CON}) respectively.  One observes that the $s$-dependence of Eq.~(\ref{eq:cross}) cancels with Eqs.~(\ref{CEN}) or (\ref{CON}), resulting in cross sections independent of the total photon-hadron energy as mentioned in Sec.~\ref{sec:intro}.

\section{Considered reactions and model inputs}
\label{sec:inputs}
\subsection{Isospin relations}
Let us first point out some trivial equalities between the cross sections of various processes related by isospin symmetry in our picture of isoscalar Pomeron exchange.
\begin{eqnarray}
 d\sigma ^{\gamma^{(*)}p \to \rho^0 \rho^0 p} &=& d\sigma ^{\gamma^{(*)}n \to \rho^0 \rho^0 n} \,,\\
  d\sigma ^{\gamma^{(*)}p \to \rho^0 \omega p} &=& d\sigma ^{\gamma^{(*)}n \to \rho^0 \omega n} \,,\\
   d\sigma ^{\gamma^{(*)}p \to \rho^0 \pi^0 p} &=& d\sigma ^{\gamma^{(*)}n \to \rho^0 \pi^0 n} \,,\\
    d\sigma ^{\gamma^{(*)}p \to \rho^0 \rho^+ n} &=& d\sigma ^{\gamma^{(*)}n \to \rho^0 \rho^- p} = 2 d\sigma ^{\gamma^{(*)}p \to \rho^0 \rho^0 p}\,,\\
    d\sigma ^{\gamma^{(*)}p \to \rho^0 \pi^+ n} &=& d\sigma ^{\gamma^{(*)}n \to \rho^0 \pi^- p} = 2 d\sigma ^{\gamma^{(*)}p \to \rho^0 \pi^0 p}\,,
\end{eqnarray}
while the isoscalar nature of the deuteron forbids the $\rho \rho $ and $\rho \pi$ channels. 

Note that the $\rho \phi$ channel cross section is expected to be very small, vanishing if the strange sea of the nucleon is symmetric (see however Refs.~\cite{Sufian:2018cpj, Liu:2019xcf}).
\subsection{Non-perturbative model inputs} 

\begin{figure}[ht]
    \centering
\includegraphics[width=0.6\textwidth]{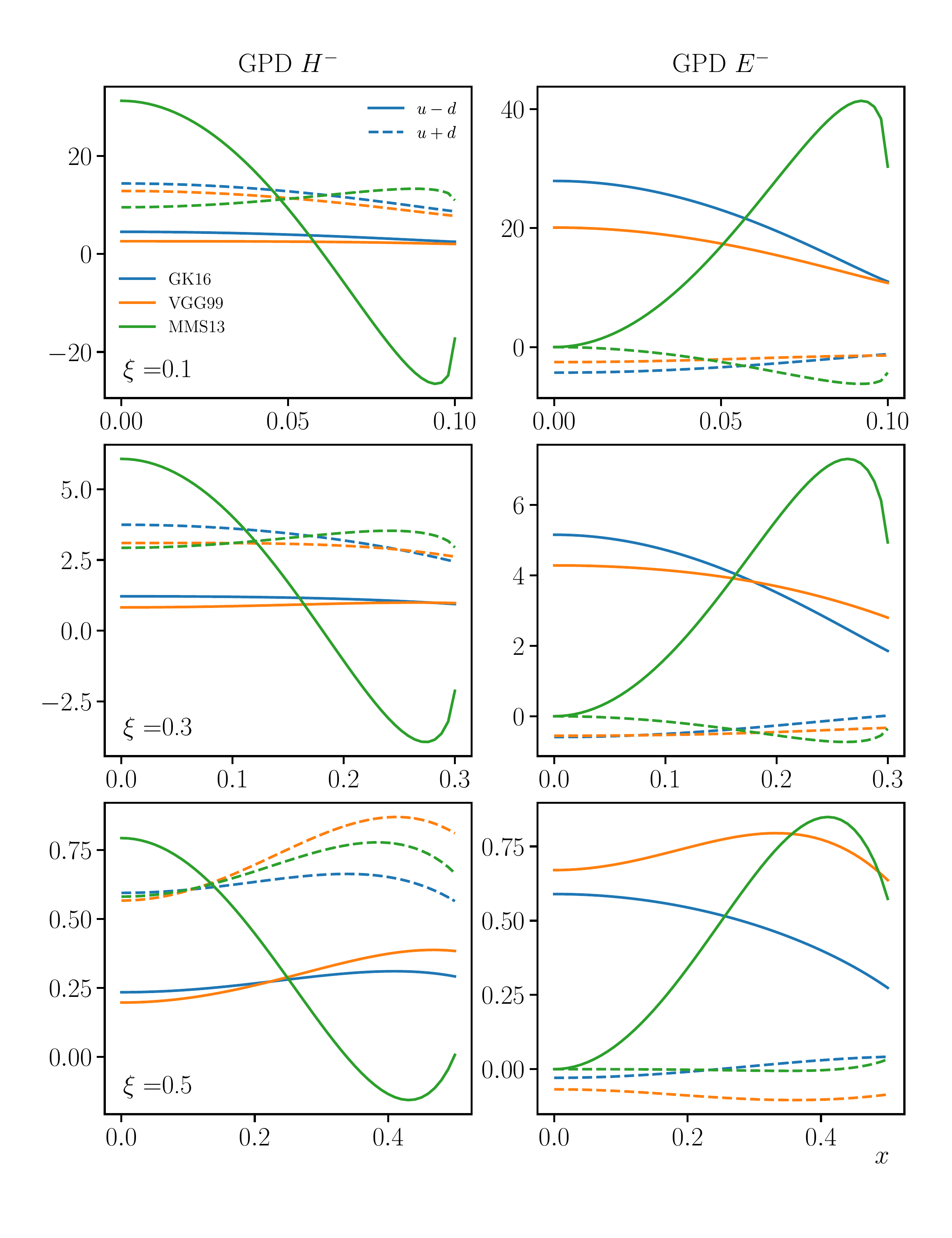}
\caption{Nucleon GPD $H^-$ (left column) and $E^-$ (right column) in the ERBL region, for isovector $u-d$ and isoscalar $u+d$ combinations in three different models~\cite{Vanderhaeghen:1999xj, Goloskokov:2007nt, Kroll:2012sm, Mezrag:2013mya}, at momentum transfer $t=t_\text{min}$. Each row has a different value of $\xi$.}
    \label{fig:gpd_nucl_vector}
\end{figure}

\begin{figure}[ht]
    \centering
\includegraphics[width=0.6\textwidth]{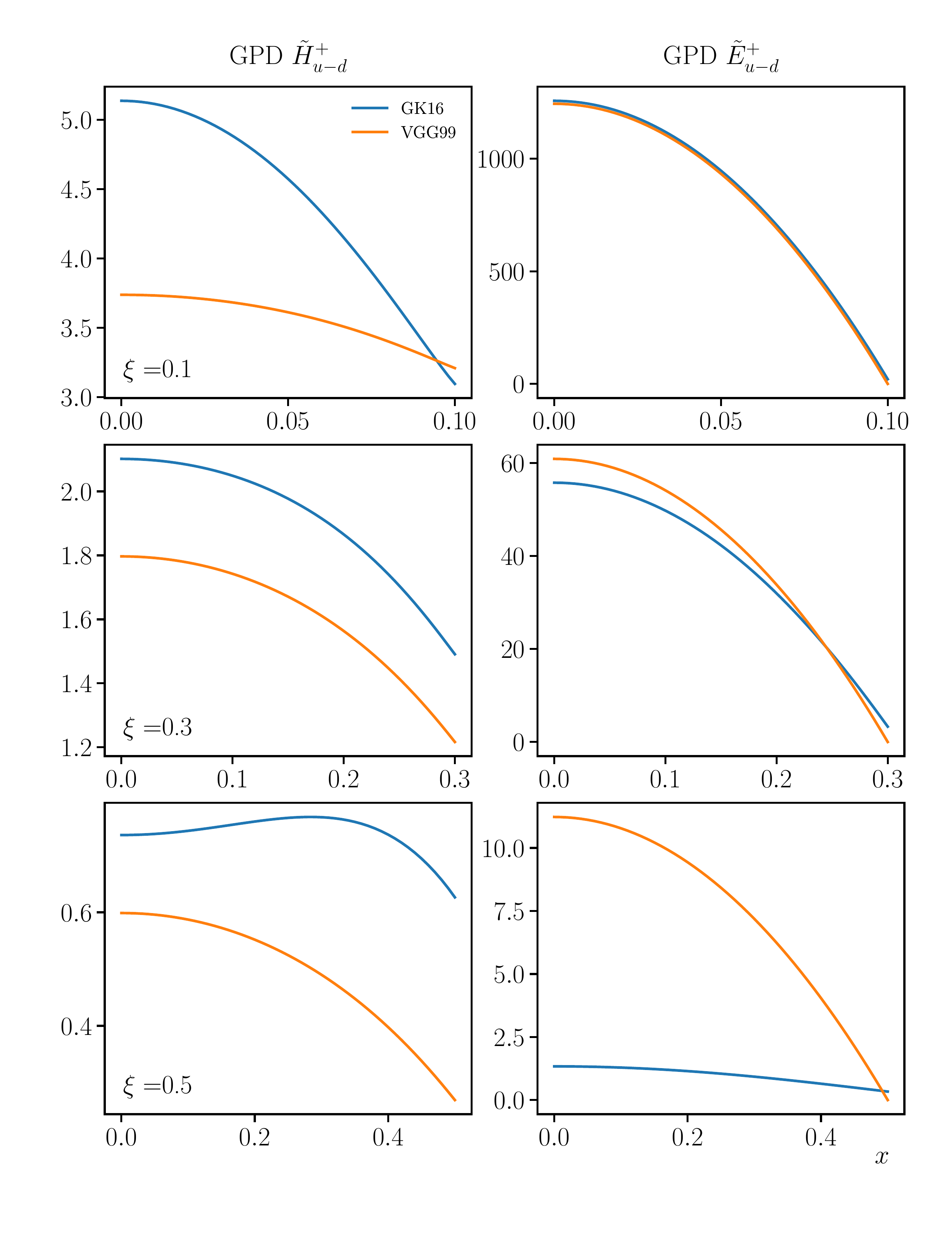}
\caption{Nucleon GPD $\widetilde{H}^+_{u-d}$ (left) and $\widetilde{E}^+_{u-d}$ (right) in the ERBL region calculated in two different models~\cite{Vanderhaeghen:1999xj, Goloskokov:2007nt, Kroll:2012sm}, at momentum transfer $t=t_\text{min}$.  Each row has a different value of $\xi$.}
    \label{fig:gpd_nucl_axial}
\end{figure}

\begin{figure}[H]
    \centering
\includegraphics[width=0.6\textwidth]{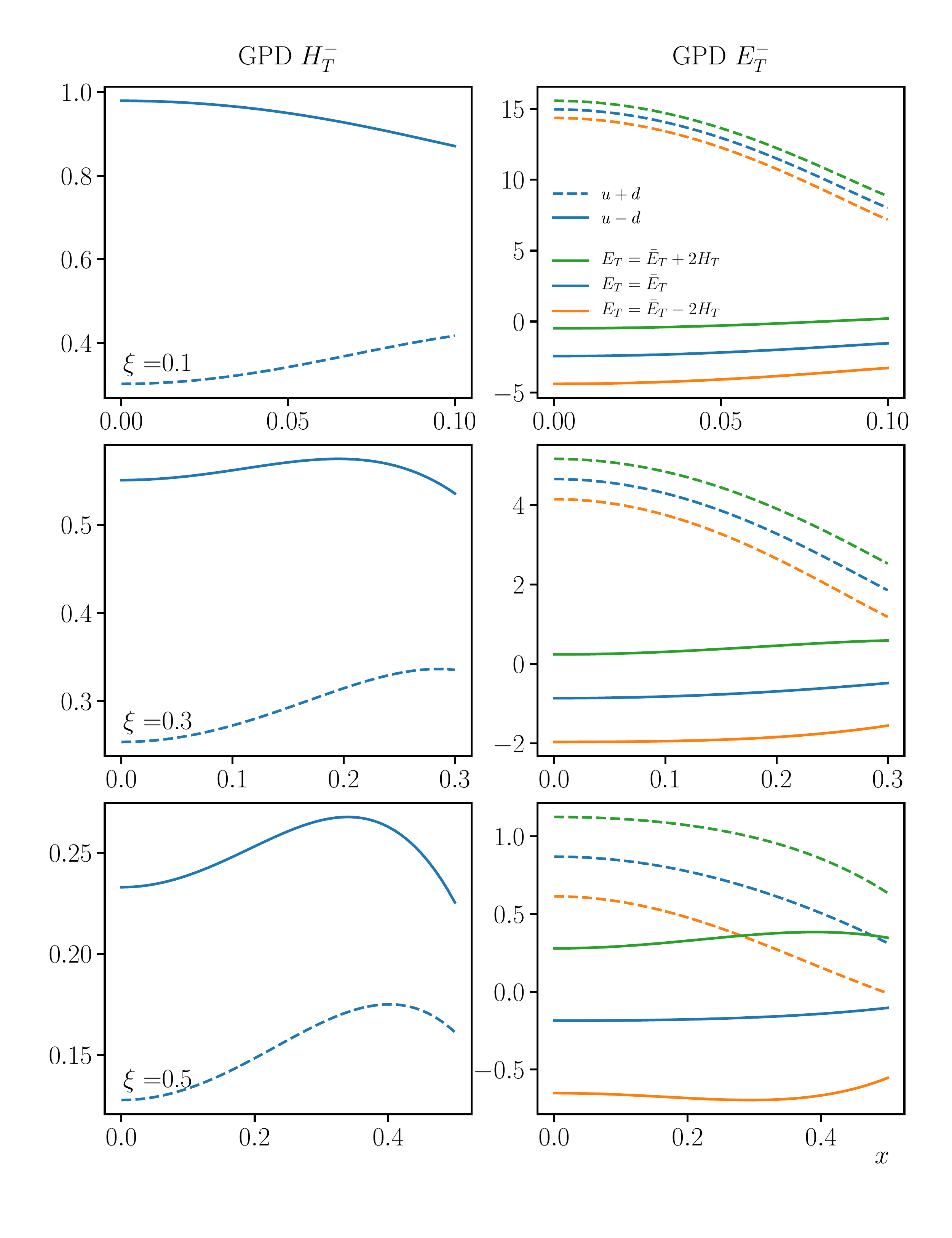}
\caption{Nucleon GPD $H_T^-$ (left column) and $E_T^-$ (right column) in the ERBL region, for isovector $u-d$ and isoscalar $u+d$, at momentum transfer $t=t_\text{min}$. Each row has a different value of $\xi$.  We consider the model of Ref.~~\cite{Goloskokov:2011rd} and use three different implementations for GPD $E_T$, see text for details.}
    \label{fig:gpd_nucl_tensor}
\end{figure}

We use different nucleon GPD models to determine the sensitivity of the observables to these badly known quantities. To guide the reader, we plot in Figs. \ref{fig:gpd_nucl_vector}, \ref{fig:gpd_nucl_axial} and \ref{fig:gpd_nucl_tensor} the vector, axial and transversity quark GPDs which enter our amplitudes. We restrict to half of the ERBL region $0\leq x \leq \xi$, since the hard amplitude selects this region of integration and the plotted charge conjugation combination of GPDs are symmetric in $x$ (see Sec.~\ref{sec:intro}). We use for the chiral-even GPDs the models VGG99~\cite{Vanderhaeghen:1999xj}, GK16~\cite{Goloskokov:2007nt, Kroll:2012sm} and MMS13~\cite{Mezrag:2013mya}  as available through the PARTONS framework~\cite{Berthou:2015oaw}. The  difference between the models, and particularly between Ref.~\cite{ Mezrag:2013mya} and the other ones will be reflected  in our estimates of the cross sections (see next section). For the chiral-odd transversity GPDs, we use the model of Ref.~\cite{Goloskokov:2011rd} for GPDs $H_T$ and $\bar{E}_T$, and three implementations discussed in Ref.~\cite{Pire:2017lfj} for GPDs $\widetilde{H}_T$ and $E_T$ (related through $\bar{E}_T \equiv E_T + 2 \widetilde{H}_T$), while we take $\widetilde{E}_T \equiv 0$.

For the chiral-even vector GPDs shown in Fig.~\ref{fig:gpd_nucl_vector}, we see clear differences between the isovector ($u-d$, which enters in $\rho$ production) and the isoscalar ($u+d$, which enters in $\omega$ production) combinations.  In the isovector case GPD $E^-$ is significantly larger than GPD $H^-$, but we have to remember that in the matrix elements entering the amplitude at $-t=(-t)_\text{min}$, GPD $E$ is multiplied by a factor of $-\xi^2/(1-\xi^2)$ compared to GPD $H$; see Eq.~(\ref{eq:corrV}).  We also notice that the MMS13 model behaves quite differently from GK16 and VGG99 for the isovector combination.  In the isoscalar case, the situation is reversed, with GPD $H^-$ clearly dominating over GPD $E^-$; all three models show similar $x$ and $\xi$ dependence.  As depicted in Fig.~\ref{fig:gpd_nucl_axial}, for the axial GPD $\widetilde{H}^+$ there are significant differences between the VGG99 and GK16 models.  GPD $\widetilde{E}^+$ dominates, but is again multiplied by the same    $\xi$-dependent factor compared to GPD $\widetilde{H}^+$ in the amplitudes; see Eq.~(\ref{eq:corrA}).  The difference between the two models at the highest value of $\xi=0.5$ is caused by the inclusion of the pion pole in the GK16 model.  Finally, in Fig.~\ref{fig:gpd_nucl_tensor} we compare the transversity GPDs $H_T^-$ and $E_T^-$, where for the isovector case we see that $H_T^-$ dominates, while for the isoscalar case $E_T^-$ does.  Again GPD $E_T$ is multiplied with the $\xi$-dependent factor in the amplitudes entering the cross sections;  see Eq.~(\ref{eq:corrT}).   The three implementations of the GK model yield relative differences that clearly grow with $\xi$.

Chiral-even and chiral-odd deuteron GPDs which are relevant for  {\em coherent} exclusive electroproduction processes as considered here, have been defined \cite{Berger:2001zb,Cosyn:2018rdm,Cosyn:2018thq} and modeled \cite{Cano:2003ju, Cosyn:2018rdm} in a convolution model including the dominant $np$ Fock state of the deuteron lightfront wave function (LFWF).  In the convolution model, the parton-deuteron helicity amplitudes are computed as a convolution of the parton-nucleon helicity amplitudes and the deuteron LFWF for the initial and final deuteron state.  Both deuteron and nucleon helicity amplitudes can be expressed as a linear combination of their respective GPDs (the operator in the helicity amplitude selecting the chiral odd or even ones), which allows to compute the deuteron GPDs; see Refs.~\cite{Cano:2003ju,Cosyn:2018rdm} for details.  By truncating the Fock expansion of the deuteron LFWF at the $np$ component, Lorentz covariance is however explicitly broken.  This has the drawback that the deuteron GPDs obtained from the convolution model do not obey the polynomiality constraints and for instance Mellin moments of these deuteron GPDs become dependent on $\xi$.  In the results for the deuteron shown later, we use the deuteron GPDs computed in this convolution model using the same nucleon GPD models discussed above.  We use the AV18~\cite{AV18} radial S- and D-wave to construct the $np$ component of the deuteron LFWF used in the convolution, previous study showed little dependence on the details of the choice of deuteron wave function~\cite{Cosyn:2018rdm}.  One kinematical feature of the convolution that is worth recalling is that the skewness entering the nucleon GPDs in the convolution ($\xi_N$) is different from that of the deuteron ($\xi$)~\cite{Cano:2003ju},
\begin{equation} \label{eq:xi_N}
\xi_N = \frac{\xi}{\frac{\alpha_N}{2}(1+\xi)-\xi}\,,
\end{equation}
 and depends on the \emph{active}\footnote{The active nucleon denotes the nucleon for which the parton correlator matrix elements are computed in the convolution.  The other non-interacting nucleon in the deuteron is then generally referred to as the \emph{spectator} nucleon. } nucleon momentum fraction $\alpha_N$ in the deuteron
 \begin{equation}
     \alpha_N = \frac{2(p_Nn)}{(p_Dn)}\,,
 \end{equation}
with $p_N$ and $p_D$ the four-momenta of the active nucleon and the deuteron.  As the deuteron LFWF is peaked at $\alpha_N\approx1$, this means $\xi_N \approx \frac{2\xi}{1-\xi}$.  This difference should be kept in mind when comparing results for the deuteron and nucleon channels.  Regarding the momentum transfer squared, we have
\begin{eqnarray}
 (-t)_\text{min} = \frac{4\xi^2 m_D^2}{1-\xi^2} \approx  \frac{4\xi_N^2 m_N^2}{1-\xi_N^2}\,,
\end{eqnarray}
with $m_D \,(m_N)$ the deuteron (nucleon) mass. As a consequence, the actual $(-t)_\text{min}$ values for a given $\xi$ and corresponding $\xi_N$ (Eq.~(\ref{eq:xi_N})) do not differ that much between the nucleon and deuteron.

    There exist many different ans\"atze for meson distribution amplitudes. We here use only two rather extreme choices to give some indication on the sensitivity of the observables to this choice: the asymptotic leading twist DA whose functional form is $\phi_{\parallel,\perp}(z) = 6z(1-z)$, and the form suggested by a holographic study \cite{Brodsky:2011xx, Forshaw:2012im, Ahmady:2016ujw} : $\phi_{\parallel,\perp}(z) = \frac{8}{\pi}\sqrt{z(1-z)}$, with their respective normalizations fixed by phenomenology \cite{Straub:2015ica}. Other models, like the dynamical chiral symmetry breaking model of Ref.~\cite{Chang:2013pq}, lead to intermediate predictions. There is no theoretical requirement that the same functional form describes $\rho_{L/T}$, $\omega_{L,T}$ and $\pi$ DAs, but for simplicity we show the results where all the meson DAs have the same form.

\section{Results}
\label{Results}

We present two different types of plots for the various reactions, where all of the points plotted were calculated at $-t=(-t)_\text{min}$.  First, we show the dependence of the cross section on the virtuality of the Pomeron $\bm{q}_{\rho}^{\,2}$ (which corresponds to the hard scale) and the photon virtuality $Q^2$.  The dependence on $\bm{q}_{\rho}^{\,2}$ is largely determined by the respective factors in Eqs.~(\ref{CEN}) and ~(\ref{CON}).  Second, to show the sensitivity of the cross sections on the model inputs discussed in Sec.~\ref{sec:inputs}, we show plots at fixed 
$\bm{q}_{\rho}^{\,2}=2~\text{GeV}^2$, $Q^2=1~\text{GeV}^2$ as a function of $\xi$.  Plots at different $\bm{q}_{\rho}^{\,2}, Q^2$ values show very similar results, with only the overall scale changing.  All plots show calculations at a renormalization and factorization scale of $\mu_R= \mu_F = 1~\text{GeV}$.

Except for Sec.~\ref{sec:electroprod}, we do not show calculations for the cross section of the electroproduction cross section
\begin{equation}
    e(p_e) + h(p_h) \to e'(p'_e) + \rho_L^0(q_\rho) + M_2(p_{M_2}) + h(p'_h) \,, 
\end{equation}
but for the (virtual) photoproduction cross sections of reaction (\ref{eq:photoreaction}).  After introducing the electron fractional energy loss
\begin{eqnarray}
 y=\frac{(qp_h)}{(p_ep_h)}\,,
\end{eqnarray}
the two cross sections are related by~\cite{Kroll:1995pv}
\begin{equation}
        \frac{\mathrm{d}\sigma^{eh}}{\mathrm{d}Q^2\mathrm{d}y\mathrm{d}\varphi \mathrm{d}\bm q^2_\rho \mathrm{d}\xi \mathrm{d}|t| }
    = \Gamma_v \left[\frac{\mathrm{d}\sigma_T}{\mathrm{d}\bm q^2_\rho \mathrm{d}\xi \mathrm{d}|t|} +\varepsilon \frac{\mathrm{d}\sigma_L}{\mathrm{d}\bm q^2_\rho \mathrm{d}\xi \mathrm{d}|t|} + \sqrt{2\varepsilon(1+\varepsilon)} \frac{\mathrm{d}\sigma_{LT}}{\mathrm{d}\bm q^2_\rho \mathrm{d}\xi \mathrm{d}|t|} \cos \varphi +\varepsilon \frac{\mathrm{d}\sigma_{TT}}{\mathrm{d}\bm q^2_\rho \mathrm{d}\xi \mathrm{d}|t|} \cos 2\varphi \right]   \,,
\label{electro}
\end{equation}
where the $\sigma_i$ are the cross sections of Eq.~(\ref{eq:cross}) with a specific photon density matrix~\cite{Kroll:1995pv}, with $\sigma_T$ corresponding to the real photoproduction cross section in the $Q^2 \rightarrow 0$ limit.
In Eq.~(\ref{electro}), the flux $\Gamma_v$ reads
\begin{equation}
   \Gamma_v = \frac{\alpha_\text{em}}{4\pi^2} \frac{y}{Q^2} \frac{1}{1-\varepsilon} \,,
\end{equation}
with $\alpha_\text{em}$ the fine-structure constant, $\varphi$ is the azimuthal angle between the electron scattering plane and the hadronic reaction plane, and $\varepsilon$ is the virtual photon linear polarization parameter,
\begin{equation}
   \varepsilon = \frac{2(1-y)-2Q^2M^2(s_{eh}-M^2)^{-2}y^2} {1+(1-y)^2+2Q^2M^2(s_{eh}-M^2)^{-2}y^2} \approx \frac{2(1-y)} {1+(1-y)^2}~~~~,~~~~ \frac{1}{1-\varepsilon}\approx \frac{1+(1-y)^2}{y^2}\,,
\end{equation}
where $s_{eh}=(p_e+p_h)^2$.
In a first step, one integrates over  $\varphi$ and restricts to the two first terms in the RHS of Eq.~(\ref{electro}). In a second step, one may study the $\varphi$ dependence to separate the longitudinal and transverse amplitudes, especially when the Rosenbluth method does not apply easily because of an unsufficient lever arm in $\varepsilon$ to deduce $\sigma_T$ and $\sigma_L$ from $\varphi$ integrated measurements at different energies.  In what follows, we show plots for the virtual photoproduction cross sections $\sigma_L$ and $\sigma_T$.

\subsection{The proton target case}
\subsubsection{Vector meson production}

\begin{figure}[ht]
    \centering
\includegraphics[width=0.6\textwidth]{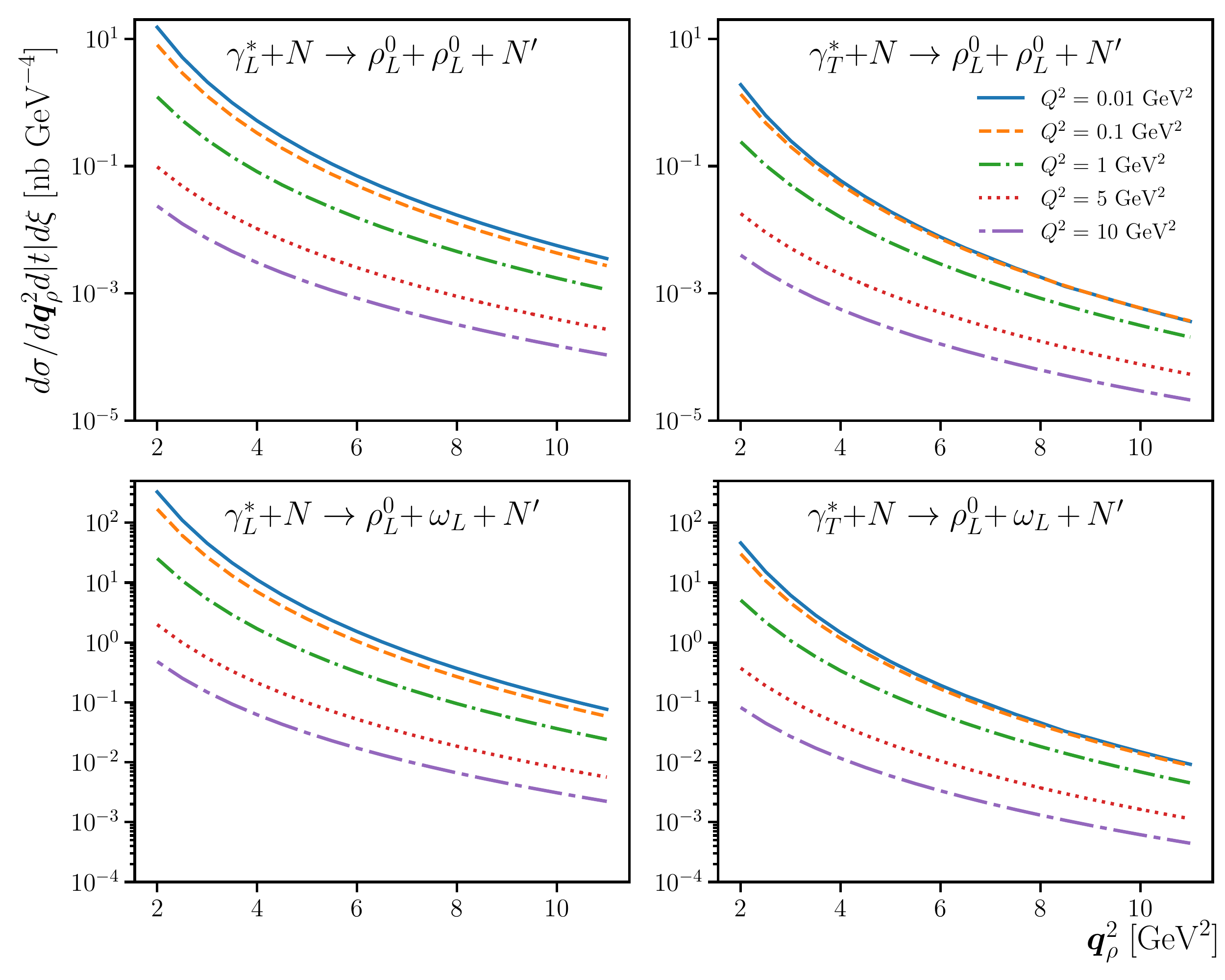}
\caption{$\bm{q}_{\rho}^{\,2}$ and $Q^2$ dependence of the cross sections of processes with a produced pair of longitudinally polarized vector mesons at $\xi=0.3$ and $-t=(-t)_\text{min}$.  This process couples to the chiral-even vector nucleon GPDs.  The GK16 model is used for the nucleon GPDs.  The upper row shows $\rho^0_L \rho^0_L$ production, which is sensitive to the isovector $u-d$ GPD. The lower row shows $ \rho^0_L \omega_L$ production, which is sensitive to the isoscalar $u+d$ GPD. }
    \label{fig:cross_pt2_rho_omega_vector}
\end{figure}

\begin{figure}[H]
    \centering
\includegraphics[width=0.6\textwidth]{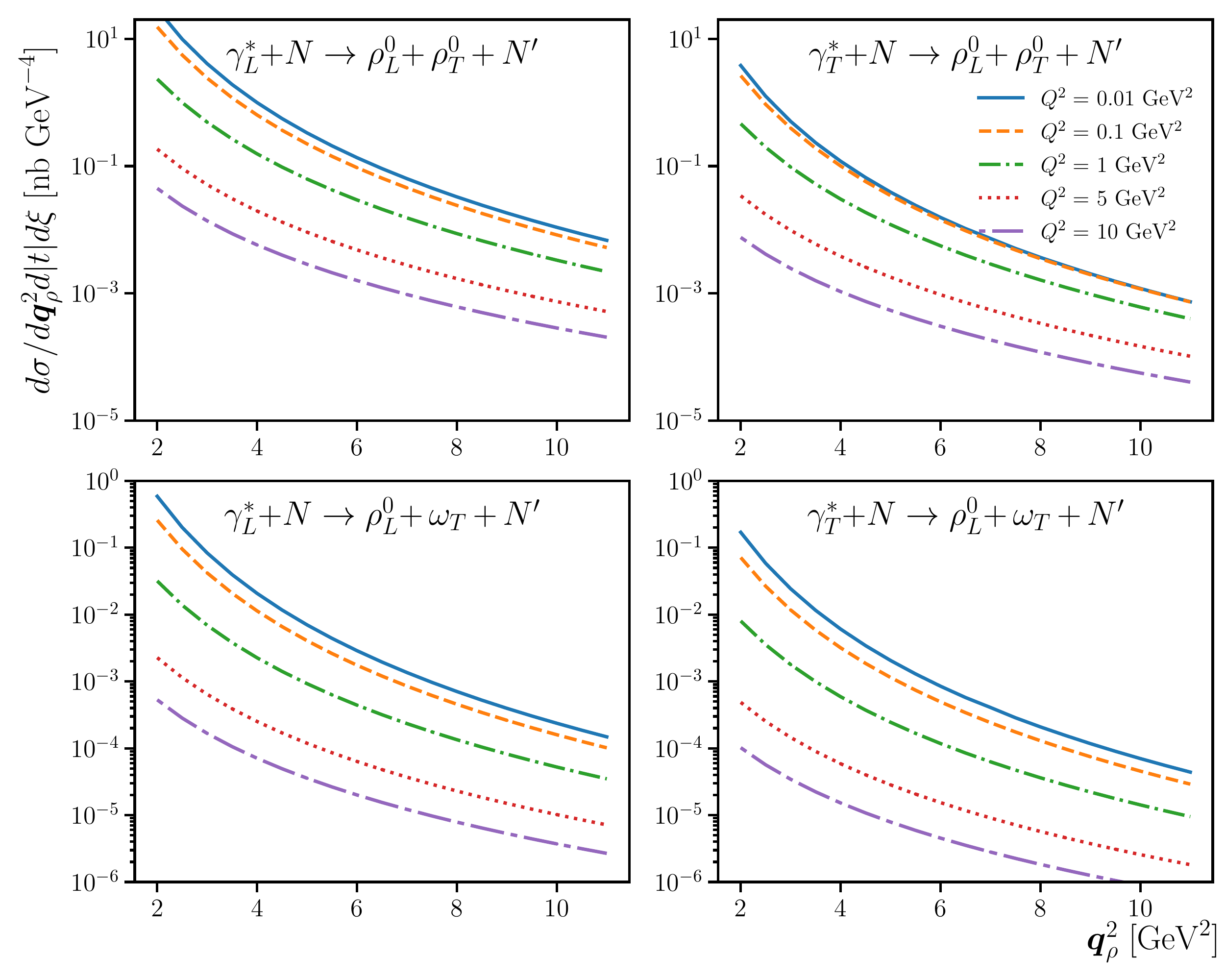}
\caption{As Fig.~\ref{fig:cross_pt2_rho_omega_vector} but for $\rho^0_L$ accompanied by a transversely polarized vector meson, which couples to the chiral-odd nucleon GPDs. 
We use the GK model for the nucleon GPDs complemented with the requirement that $E_T=\bar{E}_T$. }
    \label{fig:cross_pt2_rho_omega_tensor}
\end{figure}

The $\bm{q}_{\rho}^{\,2}$ and  $Q^2$ dependence of $\rho^0$ and $\omega$ production\footnote{To keep the notation light in this section, we mention only the $M_2$ meson of the produced pair, and leave the diffractive $\rho^0_L$ implicit.} are shown in Figs.~\ref{fig:cross_pt2_rho_omega_vector} and \ref{fig:cross_pt2_rho_omega_tensor}. One observes that the $\sigma_L$ cross sections are slightly larger than the $\sigma_T$ ones for all $Q^2$, down to small values, $\sigma_L$ still being large at $Q^2=0.01~\text{GeV}^2$.   We discuss the real photon limit in more detail in Sec.~\ref{sec:electroprod}.  In Fig.~\ref{fig:cross_pt2_rho_omega_vector}, $\omega_L$-production cross sections are larger than the $\rho_L$ ones, pointing at the dominance of GPD $H^-_{u+d}$ in the $\omega_L$-production case (see discussion of Fig.~\ref{fig:gpd_nucl_vector}).  This situation is reversed for transverse meson production in Fig.~\ref{fig:cross_pt2_rho_omega_tensor} ($\rho_T$ cross sections being larger than $\omega_T$), where the isovector $H^-_T$ dominates over the isoscalar one (see Fig.~\ref{fig:gpd_nucl_tensor}).

\begin{figure}[H]
    \centering
\includegraphics[width=0.6\textwidth]{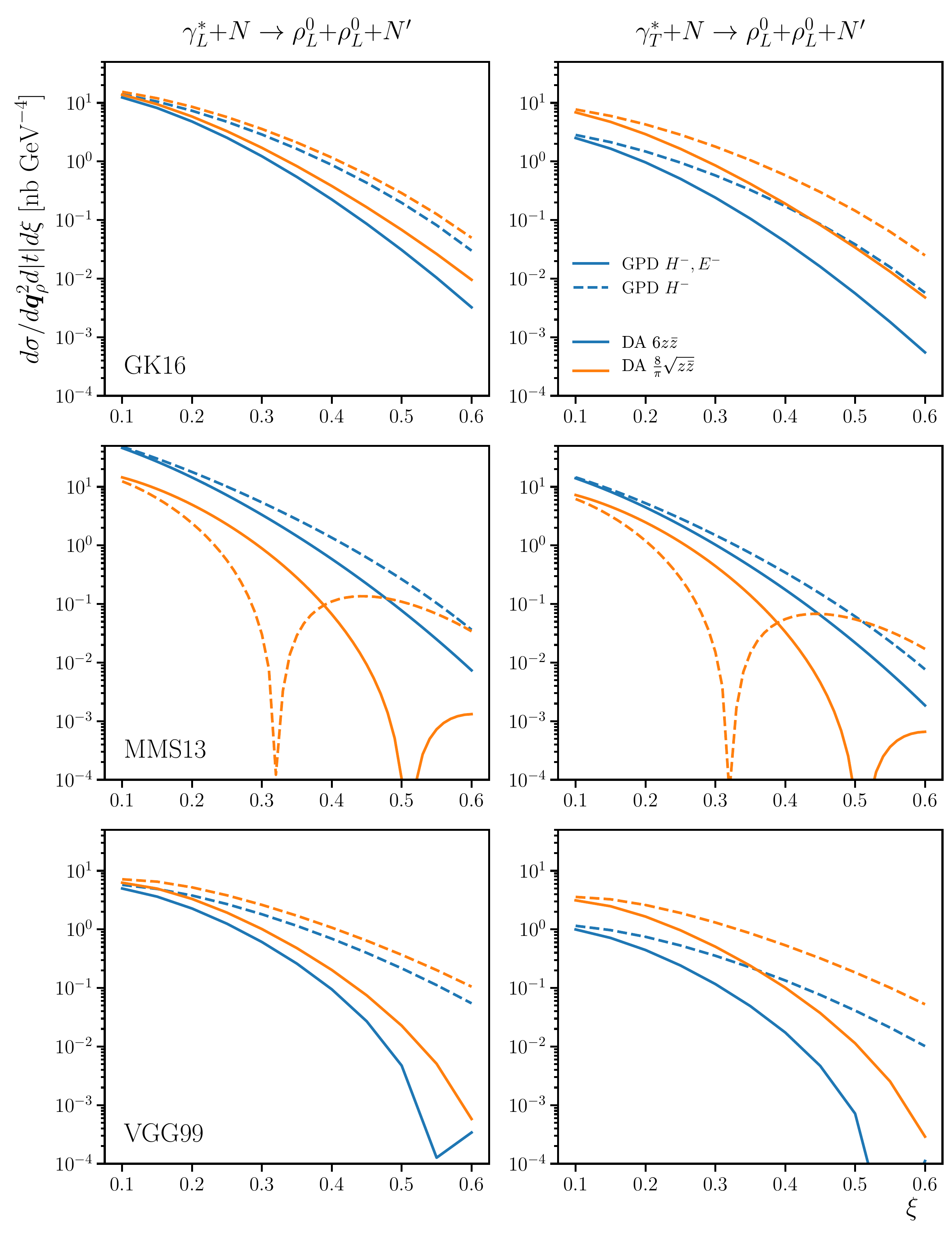}
\caption{$\xi$ dependence of the $\gamma^* + N \rightarrow \rho^0_L + \rho^0_L + N'$ cross section for $Q^2=1 ~\text{GeV}^2$, $\bm q_\rho^2=2 ~\text{GeV}^2$.  This process is sensitive to the isovector $u-d$ combination of the chiral-even vector nucleon GPDs.  We compare different GPD (rows) and DA (line color) models, see Sec.~\ref{sec:inputs} for details.  Dashed curves only include GPD $H^-$.}
    \label{fig:cross_xi_rhoL}
\end{figure}

\begin{figure}[H]
    \centering
\includegraphics[width=0.5\textwidth]{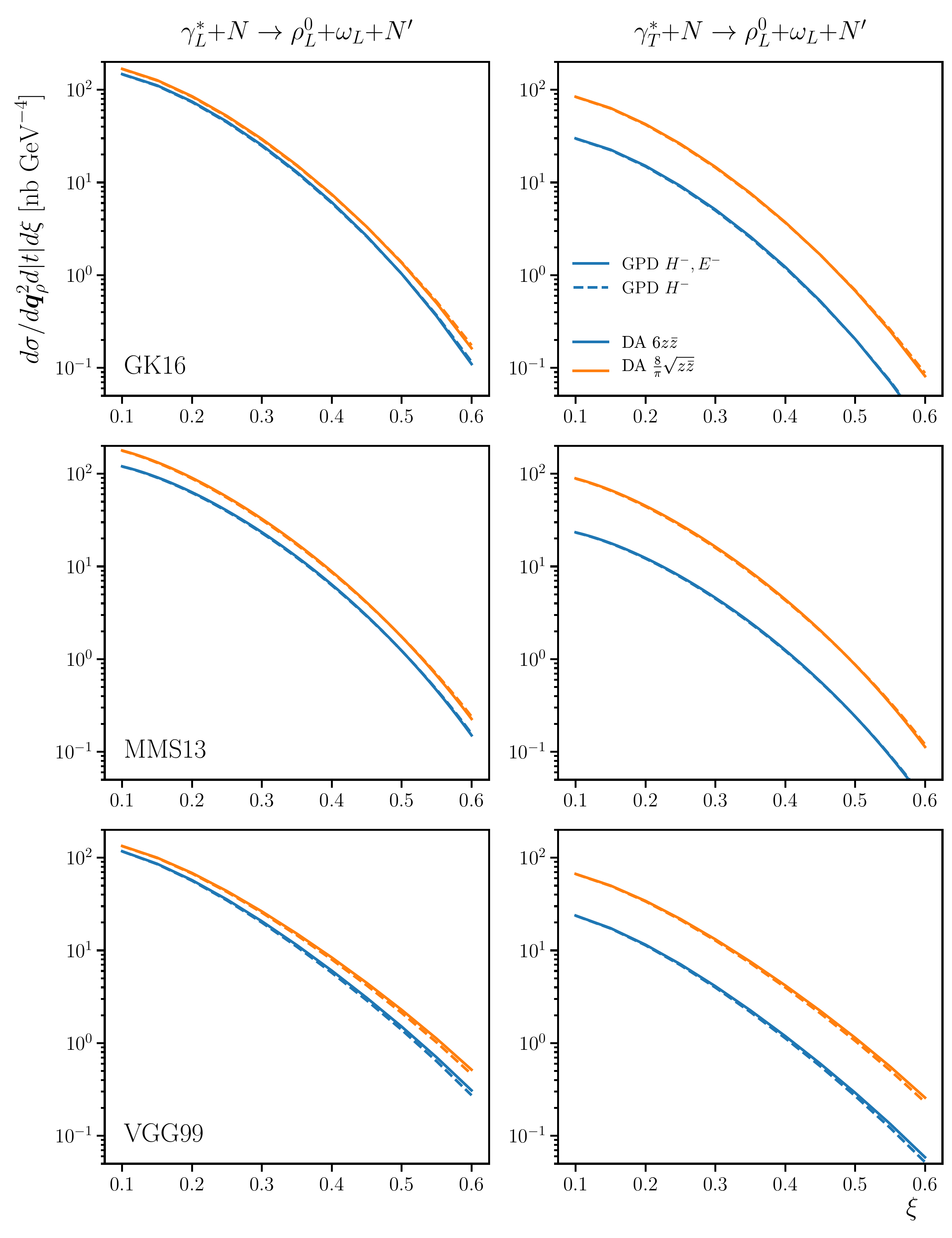}
\caption{As Fig.~\ref{fig:cross_xi_rhoL} but for $\rho_L^0 \omega_L$ production probing the isoscalar $u+d$ chiral-even vector GPDs.}
    \label{fig:cross_xi_omegaL}
\end{figure}

\begin{figure}[ht]
    \centering
\includegraphics[width=0.5\textwidth]{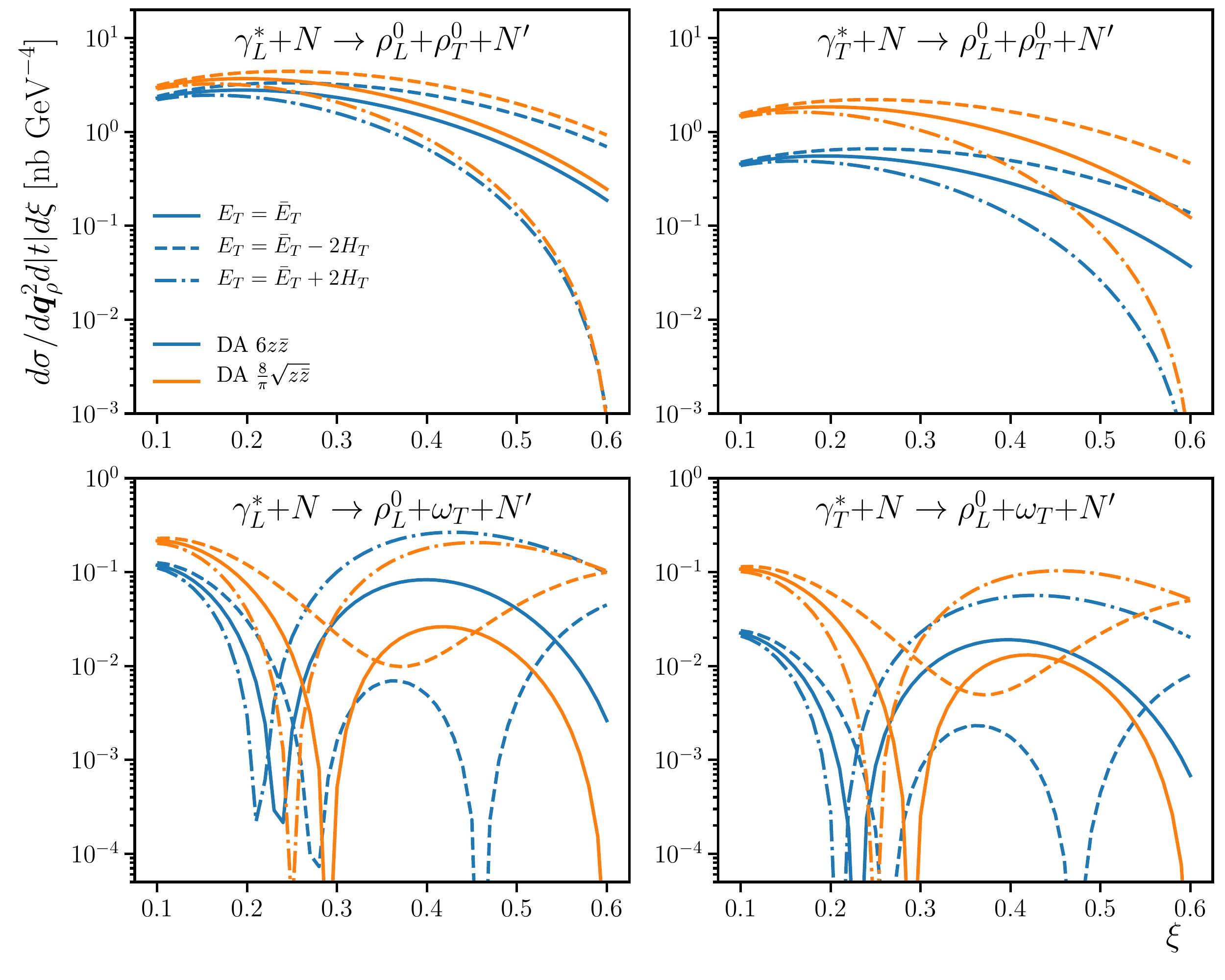}
\caption{$\xi$ dependence of the $\gamma^* + p \rightarrow \rho^0_L + (\rho^0_T/\omega_T) + p$ cross section for $Q^2=1 ~\text{GeV}^2$, $\bm q_\rho^2=2 ~\text{GeV}^2$.  This process is sensitive to the chiral-odd nucleon GPDs.  We compare different GPD and DA models, see legend and Sec.~\ref{sec:inputs} for details.}
    \label{fig:cross_xi_tensor}
\end{figure}

Next, we show the different model comparisons.  
\begin{itemize}
    \item 
For $\rho^0_L$ production, shown in Fig.~\ref{fig:cross_xi_rhoL}, one observes that the GK16 and VGG99 GPD models yield similar predictions, while the cross section with the MMS13 model behaves quite differently.  This reflects the relative differences between the $H^-,E^-$ inputs shown in Fig.~\ref{fig:gpd_nucl_vector}.  The nodes for the MMS13 cross sections can be attributed to a sign change in the nucleon amplitudes in the ERBL region, which for specific $\xi$ values yields very small cross sections.  In all three models we see that the cross section is quite sensitive to GPD $E^-$ at the larger $\xi$ values.  The two considered DAs also yield magnitudes of cross sections that are clearly separated, with the holographic DA yielding larger cross sections.  
\item For $\omega_L$ production (Fig.~\ref{fig:cross_xi_omegaL}), the differences are a lot smaller.  This again reflects the GPD inputs shown in Fig.~\ref{fig:gpd_nucl_vector}, which were quite similar for the isoscalar case.  We see GPD $E^-$  has almost no influence on the result, with $H^-$ completely dominating.  The difference between the two DAs is similar as for the $\rho_L$ case.
\item Fig.~\ref{fig:cross_xi_tensor} shows the model dependence of the $\rho^0_T$ and $\omega_T$ cross sections.  We see that for both $\rho_T^0$ and $\omega_T$ the three different implementations of the GK model are clearly separated at large $\xi$, showing sensitivity to the form of $E_T^-$. As for the longitudinal vector meson production, the holographic DA yields larger cross sections.  The $\omega_T$ cross sections show nodes for several model options, owing to a sign change in the ERBL region in the nucleon amplitudes entering the convolution. 
\end{itemize}

\subsubsection{Pseudoscalar meson production}

\begin{figure}[ht]
    \centering
\includegraphics[width=0.6\textwidth]{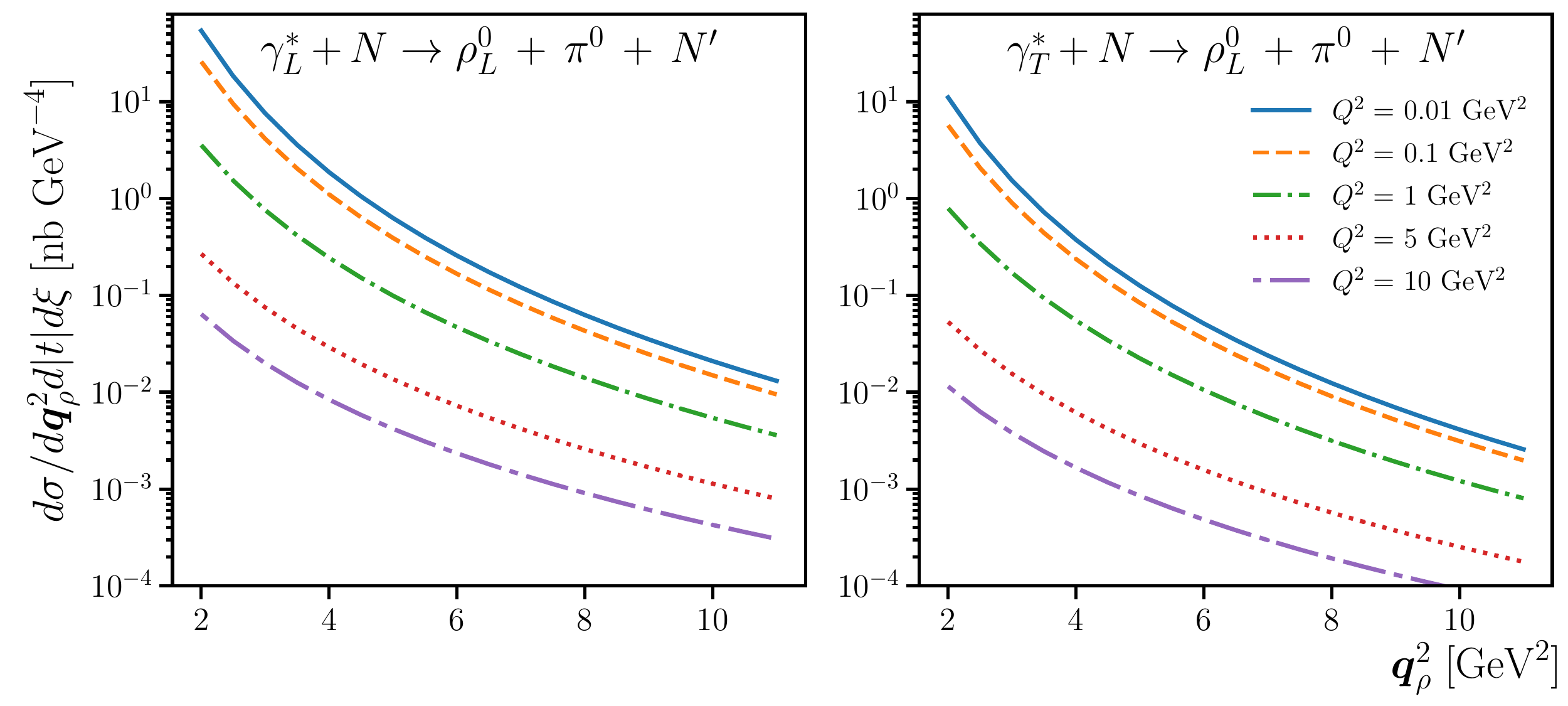}
\caption{$\bm{q}_\rho^{\,2}$ and $Q^2$ dependence of the cross sections of processes with a produced $\rho^0_L \pi^0$ at $\xi=0.3$ and $t=t_\text{min}$.  This process couples to the chiral-even axial vector nucleon GPDs.  The GK16 model is used for the axial nucleon GPDs.}
    \label{fig:cross_pt2_pi_axial}
\end{figure}

\begin{figure}[ht]
    \centering
\includegraphics[width=0.6\textwidth]{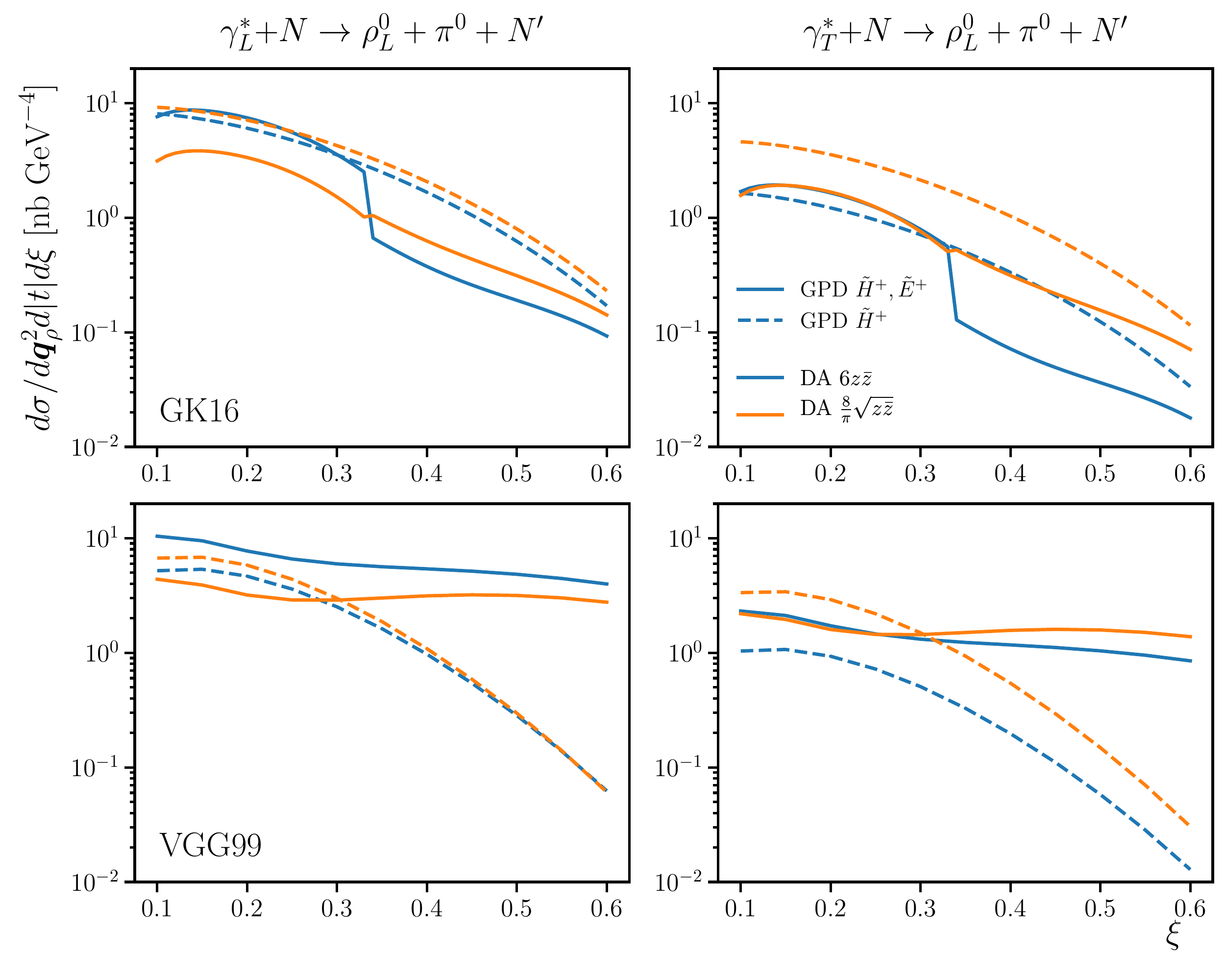}
\caption{$\xi$ dependence of the $\gamma^* + N \rightarrow \rho^0_L + \pi + N'$ cross section for $Q^2=1 ~\text{GeV}^2$, $\bm q_\rho^2=2 ~\text{GeV}^2$.  This process is sensitive to the chiral-even axial vector nucleon GPDs.  We compare different GPD (rows) and DA (line color) models. Dashed curves only include GPD $\widetilde{H}^+$.}
    \label{fig:cross_xi_pi}
\end{figure}

Fig.~\ref{fig:cross_pt2_pi_axial} shows the $\bm{q}_\rho^{\,2}$ and $Q^2$ dependence of the $\pi^0$ production cross section, with all the features discussed for two-vector meson production also present here.  The sizes of the cross sections are similar to those of $\rho^0$ production but of course depend on the magnitude of the GPD model inputs.  The different models predict a range of results, as shown in Fig.~\ref{fig:cross_xi_pi}.  In the GK16 model, the effect of the pion pole in GPD $\widetilde{E}^+$ is clearly visible around $\xi=0.34$, but the jump in the curve is much more pronounced using the asymptotic DAs.  GPD $\widetilde{E}^+$ clearly contributes at larger $\xi$ values, with especially the VGG99 results showing very little $\xi$ dependence when including both $\widetilde{H}^+$ and $\widetilde{E}^+$.  The choice of DA parameterization yields different results, with no clear trend to be observed between the effects on different GPD models. 

\subsection{The deuteron target case}
\begin{figure}[H]
    \centering
\includegraphics[width=0.6\textwidth]{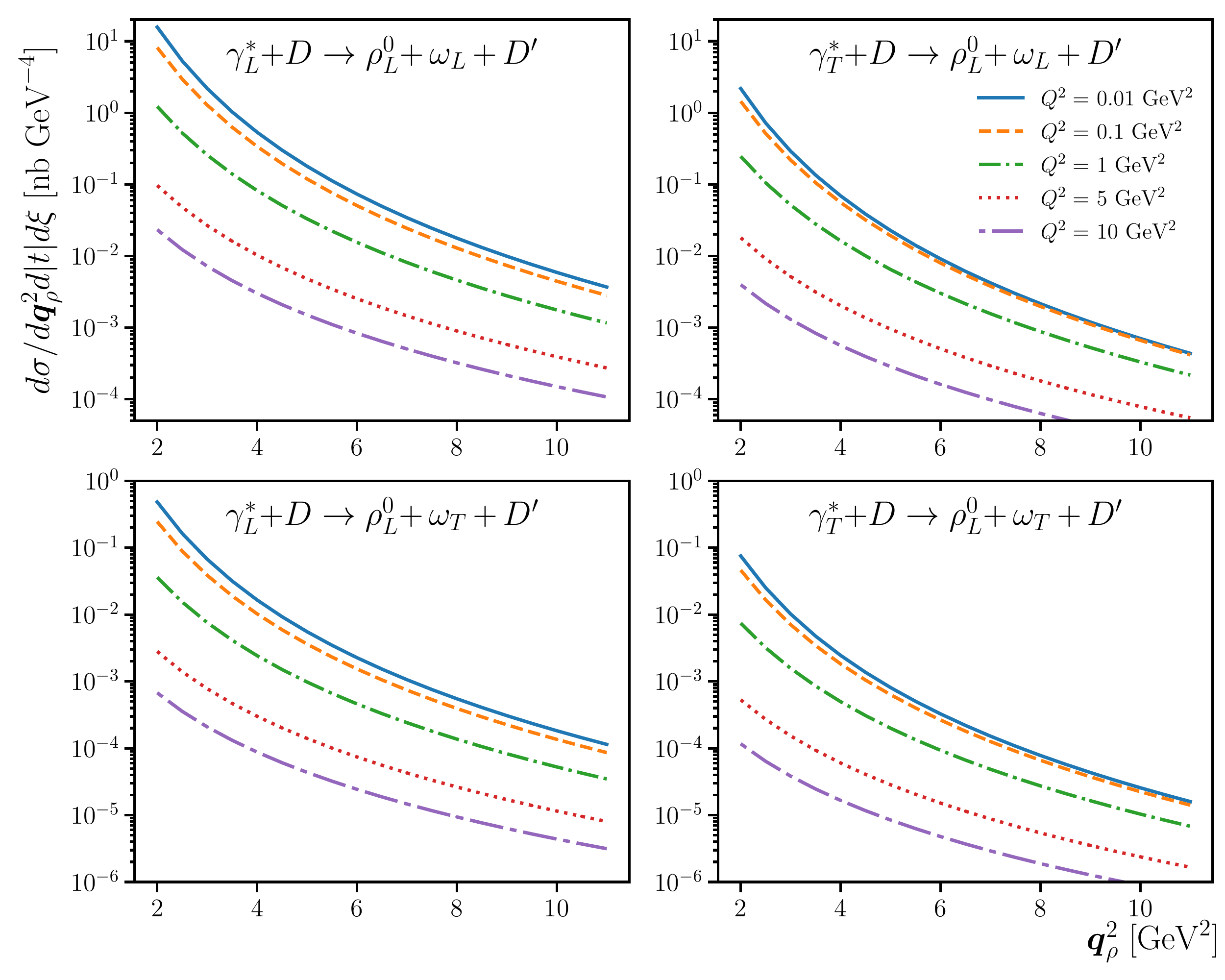}
\caption{$\bm{q}_\rho^{\,2}$ and $Q^2$ dependence of the cross-section for the coherent processes on the deuteron with a produced $\rho_L^0$ accompanied by a longitudinally or transversely  polarized $\omega$ meson at $\xi=0.15$ and $t=t_\text{min}$.  These processes couple to the chiral-even or chiral-odd isoscalar deuteron GPDs.  The GK model is used for the chiral-even and chiral-odd nucleon GPDs, with the added assumption  $E_T=\bar{E}_T$ for the latter.  The upper row shows $\omega_L$ production, which is sensitive to the chiral-even vector isoscalar deuteron GPD. The lower row shows $\omega_T$ production, which is sensitive to the chiral-odd isoscalar deuteron GPD. }
    \label{fig:cross_pt2_D_omega}
\end{figure}

\begin{figure}[H]
    \centering
\includegraphics[width=0.6\textwidth]{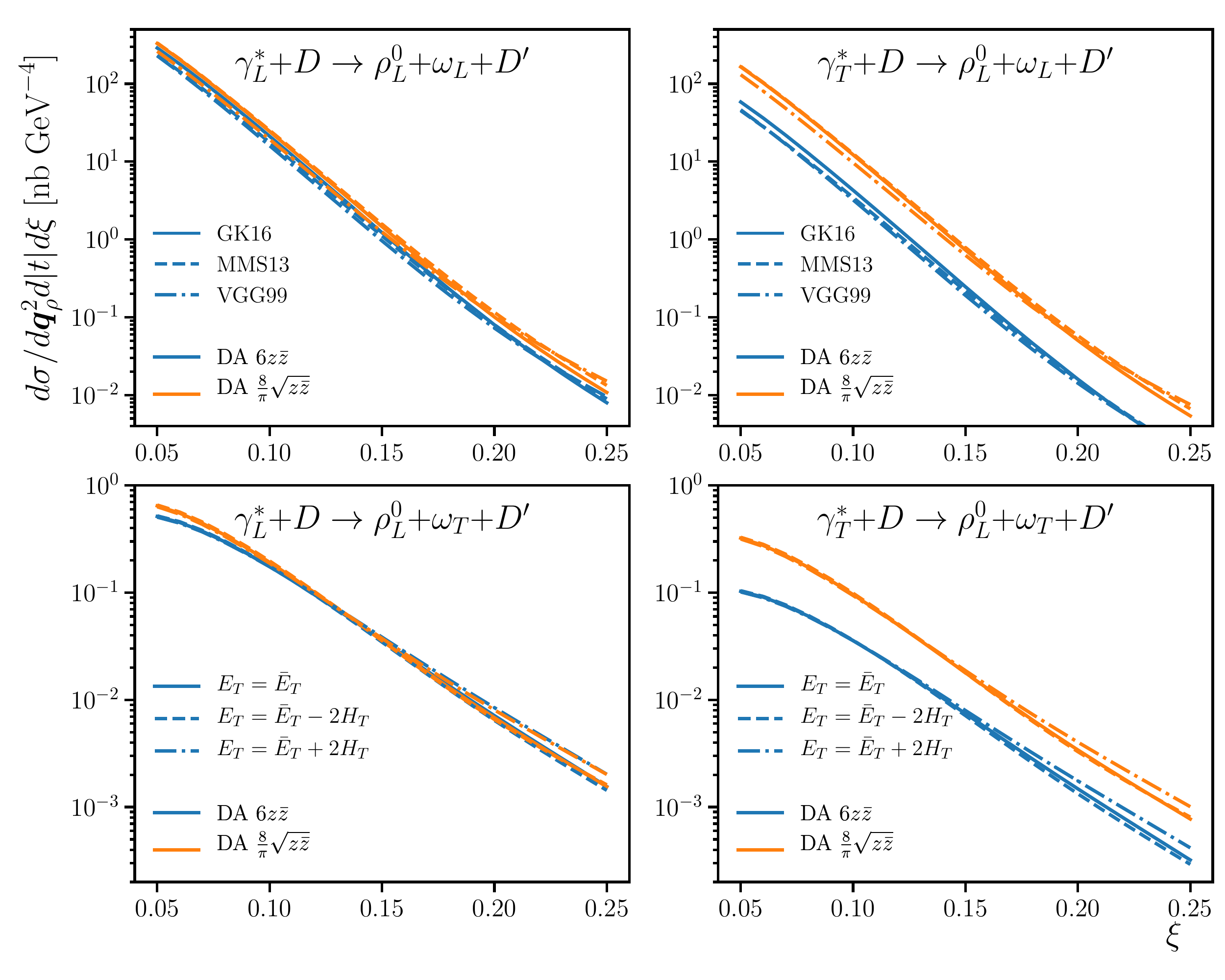}
\caption{$\xi$ dependence of the $\gamma^* + D \rightarrow \rho^0_L + \omega + D'$ coherent cross section for $Q^2=1 ~\text{GeV}^2$, $\bm q_\rho^2=2 ~\text{GeV}^2$.  This process is sensitive to the chiral-even deuteron GPDs ($\omega_L$ production, upper row) and chiral-odd deuteron GPDs ($\omega_T$ production, lower row).  We compare different GPD and DA models (see legend and text for details).}
    \label{fig:cross_xi_D}
\end{figure}
Fig.~\ref{fig:cross_pt2_D_omega} shows the $\bm{q}_\rho^{\,2}$ and $Q^2$ dependence of  coherent $\rho^0_L \omega$ production on the deuteron.  We remind that the $\rho^0_L \rho$ and $\rho^0_L \pi$ channels considered for the nucleon case are not allowed because of the isoscalar nature of the deuteron.  The trends observed in the previous reactions are present again, with $\sigma_L$ cross-sections larger than $\sigma_T$ and a similar $\bm q_\rho^2$ and $Q^2$ dependence of the curves.  Comparing $\omega$ production between the deuteron and nucleon case, we see that cross sections are smaller for the deuteron case, with the $\omega_L$ channels showing a larger drop (a bit over an order of magnitude).
Comparing the different GPD model inputs in Fig.~\ref{fig:cross_xi_D} we see very little dependence on the different models.  This can be partly attributed to the convolution, as a range of $\xi_N$ values are picked up (see Eq.~(\ref{eq:xi_N})).  A second reason, specific to the $\omega_T$ channel, was observed in the modeling of the transverse deuteron GPDs~\cite{Cosyn:2018rdm}; the deuteron transversity GPDs had little sensitivity to the implementation of the GK nucleon model (choice of $E_T$).  We do observe that the coherent deuteron cross sections drop faster with $\xi$, as can be understood by considering the average nucleon skewness values $\xi_N\approx \frac{2\xi}{1-\xi}$ entering in the convolution; see Eq.~(\ref{eq:xi_N}).

\subsection{Electroproduction}
\label{sec:electroprod}
\begin{figure}[H]
    \centering
\includegraphics[width=0.5\textwidth]{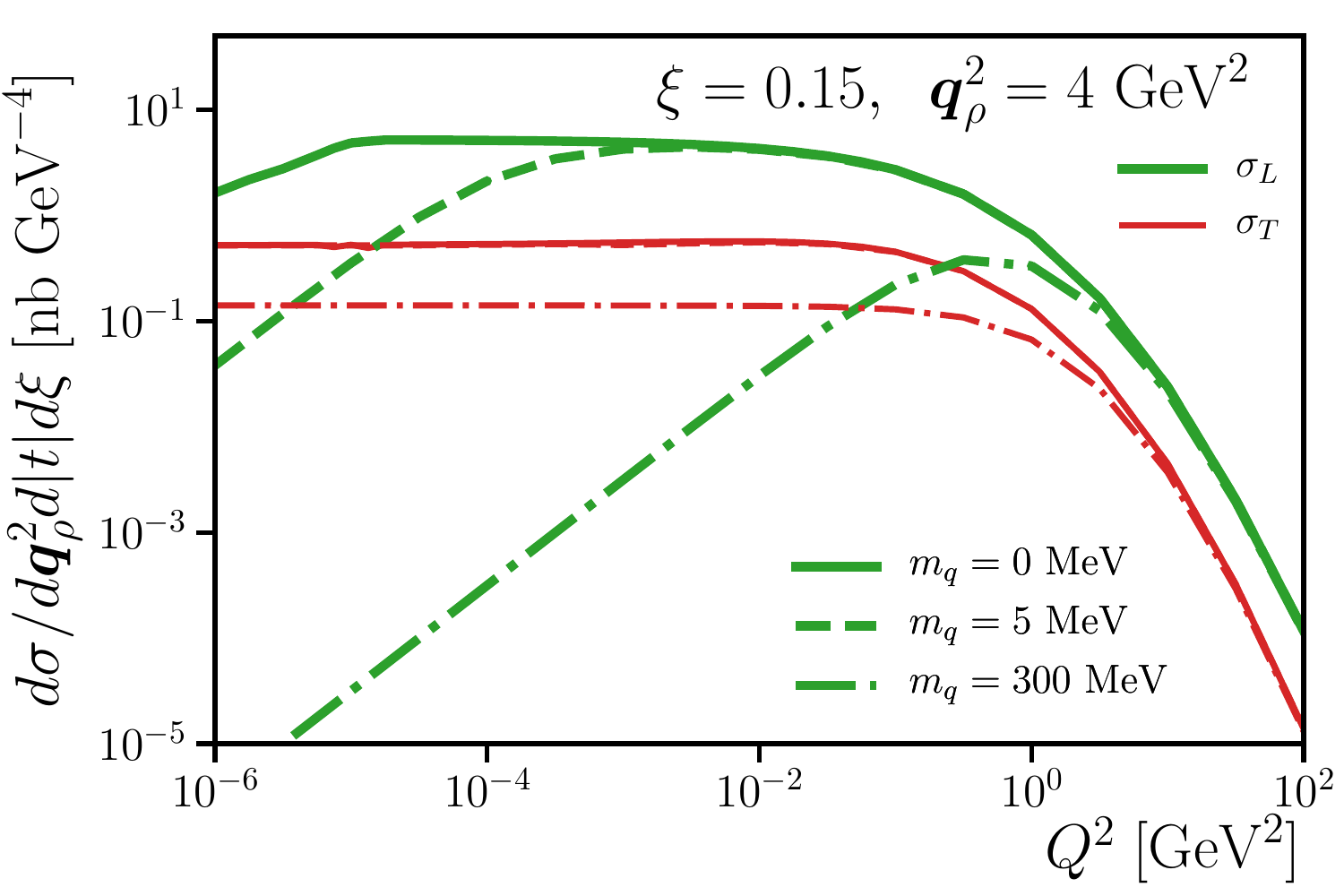}\\
\caption{  $\rho^0_L\rho^0_L$ photoproduction cross sections $\sigma_L$ and $\sigma_T$ as a function of  $Q^2$  and for different values of the quark mass $m_q$. Solid (dashed, dot-dashed) curves correspond to $m_q = 0$ ($5$, $300$) MeV.  Calculations are performed with GK16 chiral even GPDs, the asymptotic meson DAs and for $\xi=0.4$ and $\bm q_\rho^2=4~\text{GeV}^2$.}
    \label{fig:electro1}
\end{figure}

In the calculations so far we have observed large $\sigma_L$ cross sections at the lowest shown value of $Q^2=0.01~\text{GeV}^2$.  In light of this, and before we consider the quasi-real photoproduction case, we want to reassess the role of the quark mass $m_q$ in Eqs.~(\ref{P}) and (\ref{Q}), which was consistently neglected in our formalism.\footnote{Let us stress that all integrals are convergent when putting this quark mass to zero.}  Fig.~\ref{fig:electro1} shows both $\sigma_T$ and $\sigma_L$ calculated with the quark mass reinstated and at values of a current (5 MeV) and constituent (300 MeV) quark, next to the massless case.  We see that the value of $\sigma_T$ is stable for $Q^2<1~\text{GeV}$ and that the $m_q=5$ MeV curve coincides with the massless case.  For  $m_q=300$ MeV we do get a reduction of $\sigma_T$ in the quasi-real photon limit.  For $\sigma_L$, the value of $m_q$ acts as a cutoff in the denominator of Eq.~(\ref{ifgamma}) when $Q
^2$ is put to zero and determines where the $\propto Q$ behavior of the impact factor of Eq.~(\ref{ifgamma}) starts to dominate.  For $m_q=5$ MeV, we see that this happens for $Q^2$ values below roughly 0.01~GeV$^2$, while for $m_q=300$~MeV it occurs below 1~GeV$^2$.  The plot shows one choice of kinematics ($\xi=0.15, \bm q_\rho^2=4~\text{GeV}^2)$, channel ($\rho^0_L\rho^0_L$ on the nucleon), and choice of GPD (GK16) and DA (asymptotic).  Other choices yield qualitatively similar results for this figure, with only the overall scale of the calculations changing.  This overall scale can be inferred from the previous figures.

\begin{figure}[H]
    \centering
\includegraphics[width=0.5\textwidth]{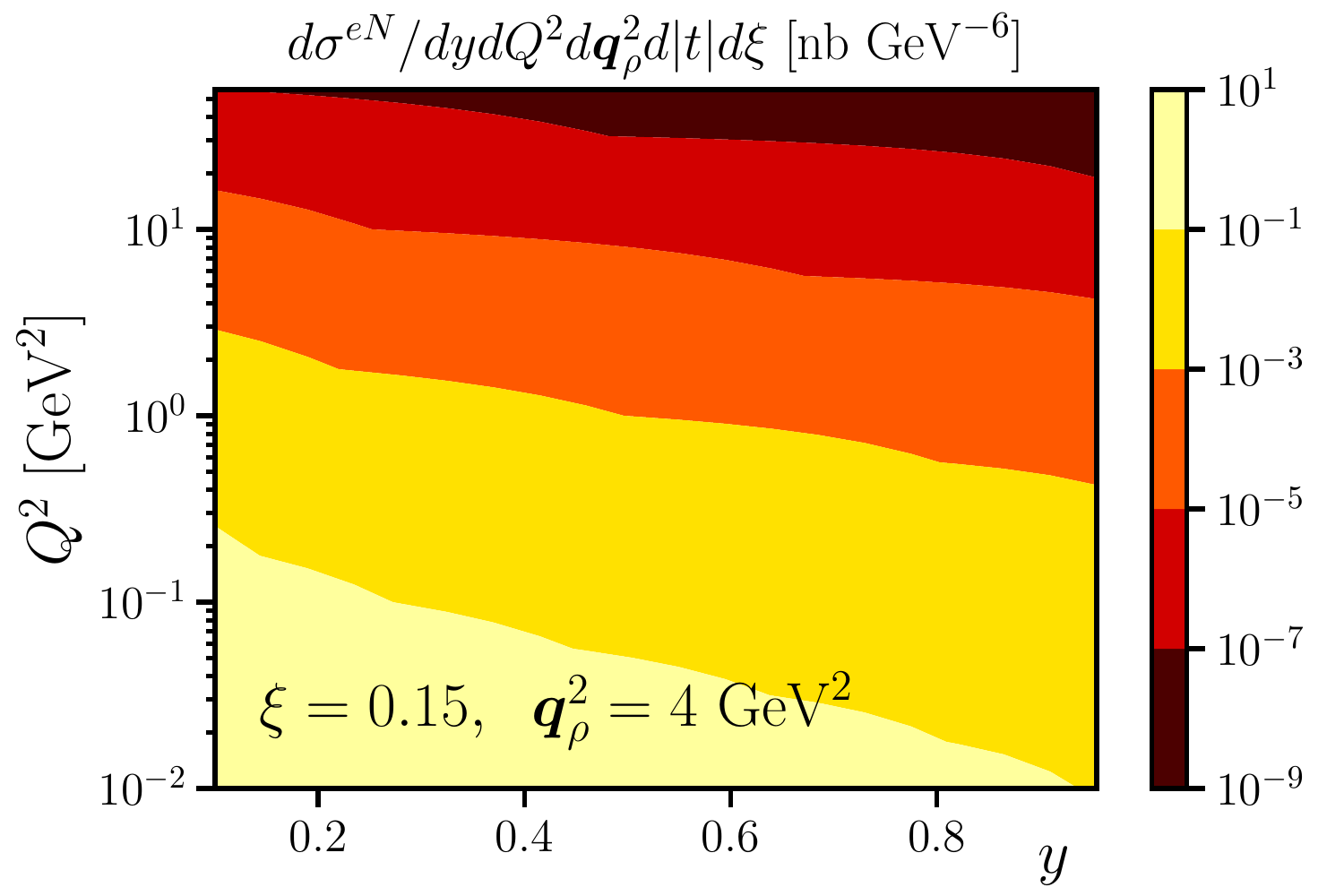}
\caption{ $\rho^0_L\rho^0_L$ electroproduction cross section on the nucleon as a function of $y$ and $Q^2$, with the GK16 chiral even GPDs, the asymptotic DA and for $\xi=0.15$ and $\bm q_\rho^2=4~\text{GeV}^2$. }
    \label{fig:electro2}
\end{figure}

To provide an estimate of the size of the electroproduction rates of our processes we show two results.
In Fig.~\ref{fig:electro2}, which shows the fully differential electroproduction cross section of Eq.~(\ref{electro}) (integrated over $\varphi$) for one characteristic kinematics, using the calculations with no quark mass. Therefore, we limit the $Q^2$ to a lower limit of 0.01~GeV$^2$.  For the quasi-real photoproduction rates, we can use the improved Weizs\"{a}cker-Williams approximation~\cite{vonWeizsacker:1934nji,Williams:1934ad,Frixione:1993yw} and write
\begin{equation}
    \frac{\mathrm{d} \sigma^{eh}}{\mathrm{d}\bm q^2_\rho \mathrm{d}\xi \mathrm{d}|t|}(Q^2_\text{max},y_1,y_2) = \int_{y_1}^{y_2} \mathrm{d}y \frac{\alpha}{2\pi} \left[ 2m_e^2 y\left( \frac{1}{Q^2_\text{max}}-\frac{1-y}{m_e^2y^2}\right)+\frac{\left((1-y)^2+1 \right)\ln \frac{Q^2_\text{max}(1-y)}{m_e^2y^2}}{y}\right]\frac{\mathrm{d} \sigma_T}{\mathrm{d}\bm q^2_\rho \mathrm{d}\xi \mathrm{d}|t|}(Q^2=0)\,,
    \label{eq:WW}
\end{equation}
 where $m_e$ is the electron mass and $\sigma_T$ is independent of $y$ as it is independent of $s$ (see Sec.~\ref{sec:intro}). Note however that since we observed that $\sigma_L$ was not negligible down to rather small values of $Q^2$, the experiment must be able to integrate over $Q^2$ with $Q^2_\text{max} \approx 10^{-2}~ \text{or}~10^{-5}~\text{GeV}^2$ (depending on the quark mass used) if we want the WW formula to be useful.  For integration limits $y_1=0.1,y_2=1.0$ in Eq.~(\ref{eq:WW}) this yields
 \begin{eqnarray}
  & \frac{\mathrm{d} \sigma_{eh}}{\mathrm{d}\bm q^2_\rho \mathrm{d}\xi \mathrm{d}|t|}= 0.0235~\,\text{nb/GeV}^{-4}\,, \qquad &\text{for}~Q^2_\text{max}=10^{-2}~\text{GeV}^2\,,\nonumber\\
  & \frac{\mathrm{d} \sigma_{eh}}{\mathrm{d}\bm q^2_\rho \mathrm{d}\xi \mathrm{d}|t|}= 0.00965~\,\text{nb/GeV}^{-4}\,, \qquad &\text{for}~Q^2_\text{max}=10^{-5}~\text{GeV}^2\,.
 \end{eqnarray}

\section{Conclusions}
This leading order analysis of the diffractive electroproduction of two mesons separated by a large rapidity gap has been demonstrated to be a promising way to access nucleon and deuteron GPDs at EIC with a particular emphasis on some very bad known features of these non-perturbative objects. Both the chiral-even and chiral-odd GPDs are entering the amplitude at the leading twist level, their contributions being well separated in an angular analysis of the $\pi \pi$ (or $\pi \pi \pi$) decay products of the $\rho$ (or $\omega$) produced in the subprocess ${\mathbb P} h \to V_{L,T} h'$. Moreover, the amplitudes have been shown to depend only on the ERBL region of the GPDs, which makes them particularly sensitive to yet unconstrained features of GPD models.  In our calculations using existing GPD parameterizations, we observed several channels where the different models could be clearly separated.  We mention the most striking examples, with the MMS13 model producing very different $\rho^0_L\rho^0_L$ cross sections compared to GK16 and VGG99, and the transversity GPD models yielding $\rho^0_L \rho^0_T$ and $\rho^0_L \omega_T$ cross sections that are clearly separated at large $\xi$.  Similarly the two different DA choices yielded cross sections that differed up to a factor of $\approx 5$.

Other channels not considered here may be interesting too. For instance the $\rho^0_L \eta$ production amplitude is sensitive to the axial gluon GPDs in the proton or deuteron, which are very poorly constrained by other channels. The $\rho^0_L \phi$ production amplitude depends mostly of the asymmetry of the strange sea in the target, which is known to be small but nevertheless does not need to vanish. We leave these topics for future studies.  

Needless to say, this leading order and leading twist study should be enlarged to include next to leading order effects, in the non-forward $\gamma^* \rho$ impact factor and  Pomeron propagator using Balitsky-Fadin-Kuraev-Lipatov (BFKL)  \cite{Fadin:1975cb, Kuraev:1977fs, Balitsky:1978ic}  techniques, and in the hard subprocess ${\mathbb P} h \to M_2 h'$ using collinear factorization techniques. While the BFKL corrections are already known for a closely related process \cite{Enberg:2003jw, Poludniowski:2003yk}, leading to quite large an enhancement  of the scattering amplitudes, a complete calculation dedicated to the process studied here needs a particular effort to be dealt with. This will take time and manpower, but should be feasible before the EIC construction is completed.  To be able to adequately measure this process, the EIC should be equipped with suitable forward detectors, which can measure the scattered nucleon or coherent deuteron down to low $-t\approx (-t)_\text{min}$ values. This is especially crucial for the coherent deuteron scattering where similar momentum transfers perpendicular to the beam as for a nucleon correspond to smaller scattering angles due to the higher momenta of the ions in the beam.

\acknowledgements
We acknowledge useful conversations with C\'edric Mezrag, Herv\'e Moutarde, Pawel Sznajder and Jakub Wagner. The work of L. S. is supported by the grant 2019/33/B/ST2/02588 of the National Science Center in Poland. This project is also co-financed by the Polish-French collaboration agreements Polonium, by the Polish National Agency for Academic Exchange and COPIN-IN2P3 and by the European Union’s Horizon 2020 research and innovation programme under grant agreement No 824093.
\bibliography{references}

\end{document}